\documentclass[preprint,showpacs,titlepage,aps,prd,tightenlines,amsmath,byrevtex,nofootinbib]{revtex4}

\usepackage{graphicx}

\begin{document}

\preprint{hep-ph/0211111}

\title{
Reactor Measurement of $\theta_{13}$ and Its Complementarity to
Long-Baseline Experiments
}

\author{H.~Minakata}
\email{E-mail: minakata@phys.metro-u.ac.jp}
\author{H.~Sugiyama}
\email{E-mail: hiroaki@phys.metro-u.ac.jp}
\author{O.~Yasuda}
\email{E-mail: yasuda@phys.metro-u.ac.jp}
\affiliation{Department of Physics, Tokyo Metropolitan University,
Hachioji, Tokyo 192-0397, Japan}

\author{K.~Inoue}
\email{E-mail: inoue@awa.tohoku.ac.jp}
\author{F.~Suekane}
\email{E-mail: suekane@awa.tohoku.ac.jp}
\affiliation{Research Center for Neutrino Science, Tohoku University,
Sendai, Miyagi, 980-8578, Japan}

\date{\today}

\vglue 1.4cm
%%%%%%%%%%%%%%%%%%%%%%%%%%%%%%%%%%%%%%%%%%%%%%%%%%%%%%%%%%%%%%%%%
%    Abstract
%%%%%%%%%%%%%%%%%%%%%%%%%%%%%%%%%%%%%%%%%%%%%%%%%%%%%%%%%%%%%%%%%
%\hfuzz=25pt
\begin{abstract}
A possibility to measure $\sin^22\theta_{13}$
using reactor neutrinos is examined in detail.
It is shown that the sensitivity
$\sin^22\theta_{13}>0.02$ can be reached with \mbox{40\,ton-year} data by
placing identical CHOOZ-like detectors at near and far distances
from a giant nuclear power plant whose total thermal energy is
\mbox{24.3\,${\text{GW}_{\text{th}}}$}.
It is emphasized that this measurement is free from
the parameter degeneracies which occur in accelerator
appearance experiments, and therefore the reactor measurement
plays a role complementary to accelerator experiments.
It is also shown that the reactor measurement may be able to
resolve the degeneracy in $\theta_{23}$
if $\sin^22\theta_{13}$ and $\cos^22\theta_{23}$
are relatively large.
\end{abstract}

\pacs{14.60.Pq,25.30.Pt,28.41.-i}
%\vskip2pc]

\maketitle

%%%%%%%%%%%%%%%%%%%%%%%%

\section{Introduction}

Despite the accumulating knowledges of neutrino masses and the
lepton flavor mixing by the atmospheric \cite{SKatm},
the solar \cite{solar,Ahmad:2002ka}, and the accelerator \cite{K2K} neutrino
experiments, the (1-3) sector of the Maki-Nakagawa-Sakata (MNS)
matrix \cite{MNS} is still in the dark.
At the moment, we only know that
$|U_{e3}| = \sin{\theta_{13}} \equiv s_{13}$ is small,
$s_{13}^2 \alt 0.03$, by the bound imposed by the CHOOZ reactor
experiment \cite{CHOOZ}.
In this paper we assume that the light neutrino sector
consists of three active neutrinos only. 
One of the challenging goals in an attempt to explore
the full structure of lepton flavor mixing would be measuring
the leptonic CP or T violating phase $\delta$ in the MNS
matrix.
If KamLAND \cite{KamLAND} confirms the Large-Mixing-Angle (LMA)
Mikheev-Smirnov-Wolfenstein (MSW) \cite{Mikheev:1986wj,Wolfenstein:1978ue}
solution of the solar neutrino problem,
the most favored one by the recent analyses of solar neutrino data
\cite{Ahmad:2002ka,solaranalysis}, we will have an open route toward the goal.
Yet, there might still exist the last impasse, namely the
possibility of too small value of $\theta_{13}$.
Thus, it is recently emphasized more and more strongly that the
crucial next step toward the goal would be the determination
of $\theta_{13}$.

In this paper, we raise the possibility that $\bar{\nu}_{e}$
disappearance experiment using reactor neutrinos could be potentially
the fastest (and the cheapest) way to detect the effects
of nonzero $\theta_{13}$.
In fact, such an experiment using the Krasnoyarsk reactor
complex has been described earlier \cite{krasnoyarsk}, in which the
sensitivity to $\sin^2{2\theta_{13}}$ can be as low as
$\sim 0.01$, an order of magnitude lower than the CHOOZ experiment.
We will also briefly outline basic features of our proposal,
and reexamine the sensitivity to $\sin^2{2\theta_{13}}$ in this paper.

It appears that the most popular way of measuring $\theta_{13}$
is the next generation long baseline (LBL) neutrino oscillation
experiments, MINOS \cite{MINOS}, OPERA \cite{OPERA}, and
the JHF phase I \cite{JHF}.
It may be followed either by conventional superbeam \cite{lowecp}
experiments, the JHF phase II \cite{JHF} and possibly others
\cite{SPL,NuMI}, or by neutrino factories
\cite{golden,nufact}.
It is pointed out, however, that the measurement
of $\theta_{13}$ in LBL experiments with only neutrino channel
(as planned in the JHF phase I) would suffer from large intrinsic
uncertainties, on top of the experimental errors, due to
the dependence on an unknown CP phase and the
sign of $\Delta m^2_{31}$ \cite{KMN02}.
Furthermore, it is noticed that the ambiguity remains in
determination of $\theta_{13}$ and other parameters even if
precise measurements of appearance probabilities in neutrino
as well as antineutrino channels are carried out, the problem of
the parameter degeneracy
\cite{FL96,Burguet-Castell:2001ez,Minakata:2001qm,Barger:2001yr,DMM02,Minakata:2002qi,KMN02}.
(For a global overview
of the parameter degeneracy, see \cite{Minakata:2002qi}.)
While some ideas toward a solution are proposed the problem is
hard to solve experimentally and it is not likely to be
resolved in the near future.

We emphasize in this paper that reactor $\bar{\nu}_{e}$
disappearance experiment provide particularly clean environment
for the measurement of $\theta_{13}$.
Namely, it can be regarded as a dedicated experiment
for determination of $\theta_{13}$; it is insensitive to
the ambiguity due to all the
remaining oscillation parameters
%(except for exceptional cases, see below)
as well as to the matter effect.
This is in sharp contrast with the features of LBL experiments
described above.
Thus, the reactor measurement of $\theta_{13}$ will provide us
valuable information complementary to the one from LBL
experiments and will play an important role in resolving
the problem of the parameter degeneracy.
It will be shown that reducing the systematic errors is crucial for
the reactor measurement of $\theta_{13}$ to be competitive in
accuracy with LBL experiments.  We will present a preliminary
analysis of its possible roles in this context.
It is then natural to think about the possibility that
one has better control by combining the two complementary way
of measuring $\theta_{13}$, the reactor and the accelerator
methods. In fact, we will show in this paper that nontrivial
relations exist between the $\theta_{13}$ measurements by both
methods thanks to the complementary nature of these two methods,
so that in the luckiest case one may be able to derive constraints
on the value of the CP violating phase $\delta$, or to determine
the neutrino mass hierarchy.

\section{Reactor experiment as a clean laboratory for
$\theta_{13}$ measurement}

Let us examine in this section how clean the measurement of
$\theta_{13}$ by the reactor experiments is.
To define our notations, we note that the standard notation 
\cite{Hagiwara:pw},
\begin{eqnarray}
U=\left[
\begin{array}{ccc}
c_{12}c_{13} & s_{12}c_{13} &   s_{13}e^{-i\delta}\\
-s_{12}c_{23}-c_{12}s_{23}s_{13}e^{i\delta} &
c_{12}c_{23}-s_{12}s_{23}s_{13}e^{i\delta} & s_{23}c_{13}\\
s_{12}s_{23}-c_{12}c_{23}s_{13}e^{i\delta} &
-c_{12}s_{23}-s_{12}c_{23}s_{13}e^{i\delta} & c_{23}c_{13}\\
\end{array}
\right],
\label{MNSmatrix}
\end{eqnarray}
is used for the MNS matrix throughout this paper 
where  $c_{ij}$ and $s_{ij}$ ($i, j = 1 - 3$) imply 
$\cos{\theta_{ij}}$ and $\sin{\theta_{ij}}$,  
respectively. The mass squared difference of neutrinos 
is defined as $\Delta m^2_{ij} \equiv m^2_i - m^2_j$ where
$m_i$ is the mass of the $i$th eigenstate.

We examine possible "contamination" by
$\delta$, the matter effect, the sign of $\Delta m^2_{31}$, and
the solar parameters one by one.
We first note that, due to its low neutrino energy of
a few MeV, the reactor experiments are inherently
disappearance experiments, which can measure only the survival
probability $P(\bar{\nu}_{e} \rightarrow \bar{\nu}_{e})$.
It is well known that the survival probability does not
depend on the CP phase $\delta$ in arbitrary matter densities
\cite{delta-indep}.

In any reactor experiment on the Earth, short or long baseline,
the matter effect is very small because the energy is quite low
and can be ignored to a good approximation.
It can be seen by comparing the matter and the vacuum effects
(as the matter correction comes in only through this combination
in the approximate formula in \cite{golden})
\begin{eqnarray}
\frac{aL}{|\Delta_{31}|} = 2.8 \times 10^{-4}
\left(\frac{|\Delta m^2_{31}|}{2.5 \times10^{-3}\,\mbox{eV}^2} \right)^{-1}
\left(\frac{E}{4\,\mbox{MeV}} \right)
\left(\frac{\rho}{2.3\,\mbox{g}\cdot\mbox{cm}^{-3}} \right)
\left(\frac{Y_e}{0.5} \right),
\end{eqnarray}
where
\begin{eqnarray}
\Delta_{ij} \equiv \frac{\Delta m^2_{ij} L}{2E}
\end{eqnarray}
with $E$ being the neutrino energy and $L$ baseline length.
The best fit value of $|\Delta m^2_{31}|$ is given by
$|\Delta m^2_{31}|=2.5\times10^{-3}\,\mbox{eV}^2$ from the
Super-Kamiokande atmospheric neutrino
data \cite{shiozawa},
and we use this as the reference value for
$|\Delta m^2_{31}|$ throughout this paper.
$a = \sqrt{2} G_F N_e$ denotes the index of refraction
in matter with $G_F$ being the Fermi constant and
$N_e$ the electron number density in the Earth which is related
to the Earth matter density $\rho$ as
$N_e = Y_e\rho/m_p$ where $Y_e$ is proton fraction.
Once we know that the matter effect is negligible we
immediately recognize that the survival probability
is independent of the sign of $\Delta m^2_{31}$.

Therefore, the vacuum probability formula applies.
The general probability formula in vacuum
is analytically written as \cite{Hagiwara:pw}
\begin{eqnarray}
\left\{ \begin{array}{c}
P(\nu_{\alpha} \rightarrow \nu_{\beta})\\
P(\bar{\nu}_{\alpha} \rightarrow\bar{\nu}_{\beta})
\end{array}\right\}
&=&\delta_{\alpha\beta}
-4\sum_{j<k}\mbox{\rm Re}\left(U_{\alpha j}U_{\beta j}^\ast
U_{\alpha k}^\ast U_{\beta k}\right)\sin^2\left(
{\Delta m^2_{jk}L \over 4E}\right)\nonumber\\
&{\ }&\mp2\sum_{j<k}\mbox{\rm Im}\left(U_{\alpha j}U_{\beta j}^\ast
U_{\alpha k}^\ast U_{\beta k}\right)\sin\left(
{\Delta m^2_{jk}L \over 2E}\right),
\label{Pgeneral}
\end{eqnarray}
where $\alpha, \beta = e, \mu, \tau$, and
the minus and signs in front of
the $\mbox{\rm Im}\left(U_{\alpha j}U_{\beta j}^\ast
U_{\alpha k}^\ast U_{\beta k}\right)$
term in (\ref{Pgeneral}) correspond to neutrino and
antineutrino channels, respectively.
From (\ref{Pgeneral}) the exact expression for
$P(\bar{\nu}_{e} \rightarrow \bar{\nu}_{e})$
is given by
\begin{eqnarray}
\hspace*{-95mm}1-P(\bar{\nu}_{e} \rightarrow \bar{\nu}_{e}) &=&
4\sum_{j<k}|U_{ej}|^2|U_{ek}|^2\sin^2\left(
{\Delta m^2_{jk}L \over 4E}\right)\nonumber \\
\hspace*{-15mm}
&=&\sin^2{2 \theta_{13}}
\sin^2{\frac{\Delta_{31}}{2}}
+ \frac{1}{\,2\,} c^2_{12}
\sin^2{2 \theta_{13}}
\sin{\Delta_{31}}\sin{\Delta_{21}} 
\nonumber \\
&+&
\left(
c^4_{13} \sin^2{2 \theta_{12}} +
c^2_{12} \sin^2{2 \theta_{13} \cos{\Delta_{31}}}
\right)
\sin^2{\frac{\Delta_{21}}{2}},
\label{Pvac}
\end{eqnarray}
where the parametrization (\ref{MNSmatrix}) has been used in the second
line.
The three terms in the second line of (\ref{Pvac}) are suppressed
relative to the main depletion term,
the first term of the right-hand-side of (\ref{Pvac}) , by
$\epsilon$,
$\epsilon^2/\sin^2{2 \theta_{13}}$,
$\epsilon^2$, respectively, where
$\epsilon \equiv \Delta m^2_{21}/|\Delta m^2_{31}|$.
Assuming that
$|\Delta m^2_{31}| = (1.6\mbox{-}3.9) \times 10^{-3}\,\mbox{eV}^2$
\cite{shiozawa},
$\epsilon \simeq 0.1\mbox{-}0.01$ for the LMA MSW
solar neutrino solution \cite{Ahmad:2002ka,solaranalysis}.
Then, the first and the third terms in the second line can be
ignored, although the second term can be of order unity compared
with the main depletion term provided that $\epsilon \simeq 0.1$.
(Notice that we are considering the measurement of
$\sin^2{2 \theta_{13}}$ in the range of 0.1-0.01.)
Therefore, assuming that $|\Delta m^2_{31}|$ is determined by LBL
experiments with good accuracy, the reactor $\bar{\nu}_e$
disappearance experiment gives us a clean measurement of $\theta_{13}$
which is independent of any solar parameters except for
the case of high $\Delta m^2_{21}$ LMA solutions.

If the high $\Delta m^2_{21}$ LMA solution with
$\Delta m^2_{21} \sim 10^{-4}\,\mbox{eV}^2$ turns out to be
the right one,
we need a special care for the second term of the second line of
(\ref{Pvac}). In this case, the determination of $\theta_{13}$ and
the solar angle $\theta_{12}$ is inherently coupled,\footnote{
The effect of nonzero $\theta_{13}$ for measurement of
$\theta_{12}$ at KamLAND is discussed in \cite{concha-carlos}.}
and we would need joint analysis of near-far detector complex
(see the next section) and KamLAND.

\section{Near-far detector complex:
basic concepts and estimation of sensitivity}

In order to obtain good sensitivity to $\sin^2{2 \theta_{13}}$,
selection of an optimized baseline and having the small statistical
and systematic errors are crucial.
For instance, the baseline length that gives the oscillation maximum
for reactor $\bar{\nu}_e$'s which have typical energy \mbox{4\,MeV}
is \mbox{1.7\,km} for
$\Delta m^2 \simeq 2.5\times 10^{-3}\,\mbox{eV}^2$.
Along with this baseline selection, if systematic
and statistical errors can be reduced to \mbox{1\,\%} level, which is
2.8 times better than the CHOOZ experiment \cite{CHOOZ},
an order of magnitude improvement for the
$\sin^2{2 \theta_{13}}$ sensitivity is possible at
$\Delta m^2 \simeq 2.5\times 10^{-3}\,\mbox{eV}^2$.
In this section we demonstrate that such kind of experiment is
potentially possible if we place a CHOOZ-like detector at a
baseline \mbox{1.7\,km} in \mbox{200\,m} underground
near a reactor of $24.3\,{\text{GW}_{\text{th}}}$ thermal power.
 The reactor can be regarded as a simplified one
of the Kashiwazaki-Kariwa nuclear power plant
which consists of seven reactors and
whose maximum energy generation is $24.3\,{\text{GW}_{\text{th}}}$.

Major part of systematic errors is caused by uncertainties of the
neutrino flux calculation, number of protons,
and the detection efficiency. For instance,
in the CHOOZ experiment,
the uncertainty of the neutrino flux is \mbox{2.1\,\%}, that of
number of protons is \mbox{0.8\,\%}, and that of detection efficiency
\mbox{1.5\,\%} as is shown in the Table \ref{table}.
The uncertainty of the neutrino flux includes ambiguities of the
reactor thermal power generation, the reactor fuel component,
the neutrino spectra from fissions, and so on.
The uncertainty of the detection efficiency includes
systematic shift of defining the fiducial volume.
These systematic uncertainties, however, cancel out
if identical detectors are placed near and far from the reactors and data
taken at the detectors are
compared.\footnote{This is more or less the strategy taken in the
Bugey experiment \cite{bugey}.  The Krasnoyarsk group also plans in
their Kr2Det proposal \cite {krasnoyarsk} to construct two identical
\mbox{50\,ton} liquid scintillators at \mbox{1100\,m}
and \mbox{150\,m} away from the
Krasnoyarsk reactor. They indicate that the systematic error can be
reduced to \mbox{0.5\,\%} by comparing the front and far detector.}

To estimate how good the cancellation will be, we study the case
of the Bugey experiment, which uses three identical detectors
to detect reactor neutrinos at \mbox{14/40/90\,m}.
For the Bugey case, the uncertainty of the neutrino flux
improved from \mbox{3.5\,\%} to \mbox{1.7\,\%}
and the error on the solid angle
remained the same ($0.5\,\% \rightarrow 0.5\,\%$).
If each ratio of the improvement for the Bugey case is directly
applicable to our case, the systematic uncertainty will
improve from \mbox{2.7\,\%} to \mbox{0.8\,\%} as shown
in the Table \ref{table}.
The ambiguity of the solid angle will be
negligibly small because the absolute baseline is much longer than
the Bugey case.
 We are thinking of a case that a front detector is located at \mbox{300\,m}
away from the reactor we consider.
 In the actual setting with the Kashiwazaki-Kariwa power plant
two near detectors may be necessary due to extended array of
seven reactors.
Hereafter, we take \mbox{2\,\%} and \mbox{0.8\,\%}
as the reference values of the relative systematic error
$\sigma_{\text{sys}}$
for the total number of $\bar{\nu}_e$ events in our analysis.
Let us examine the physics potential of such a reactor experiment
assuming these reference values for the systematic error.
We take, for concreteness, the Kashiwazaki-Kariwa reactor of
$24.3\,{\text{GW}_{\text{th}}}$ thermal power and assume
its operation with \mbox{80\,\%} efficiency.
Two identical liquid scintillation detectors are located
at \mbox{300\,m} and \mbox{1.7\,km} away from the reactor
and assumed to detect $\bar{\nu}_e$ by
delayed coincidence with \mbox{70\,\%} detection efficiency.
The $\bar{\nu}_e$'s of \mbox{1-8\,MeV} visible energy,
$E_{\text{visi}} = E_{\bar{\nu}_e} - 0.8\,{\text{MeV}}$, are used
and the number of events are counted in 14 bins of \mbox{0.5\,MeV}\@.
Without oscillation,
%
%20\,ton-year measurement yields 40,000 $\bar{\nu}_e$ events
%
a \mbox{10 (40)\,ton-year} measurement at the far detector
yields 20,000 (80,000) $\bar{\nu}_e$ events
which is naively comparable to
a \mbox{0.7 (0.35)\,\%} statistical error.

\begin{figure}[h]
\includegraphics[scale=0.8]{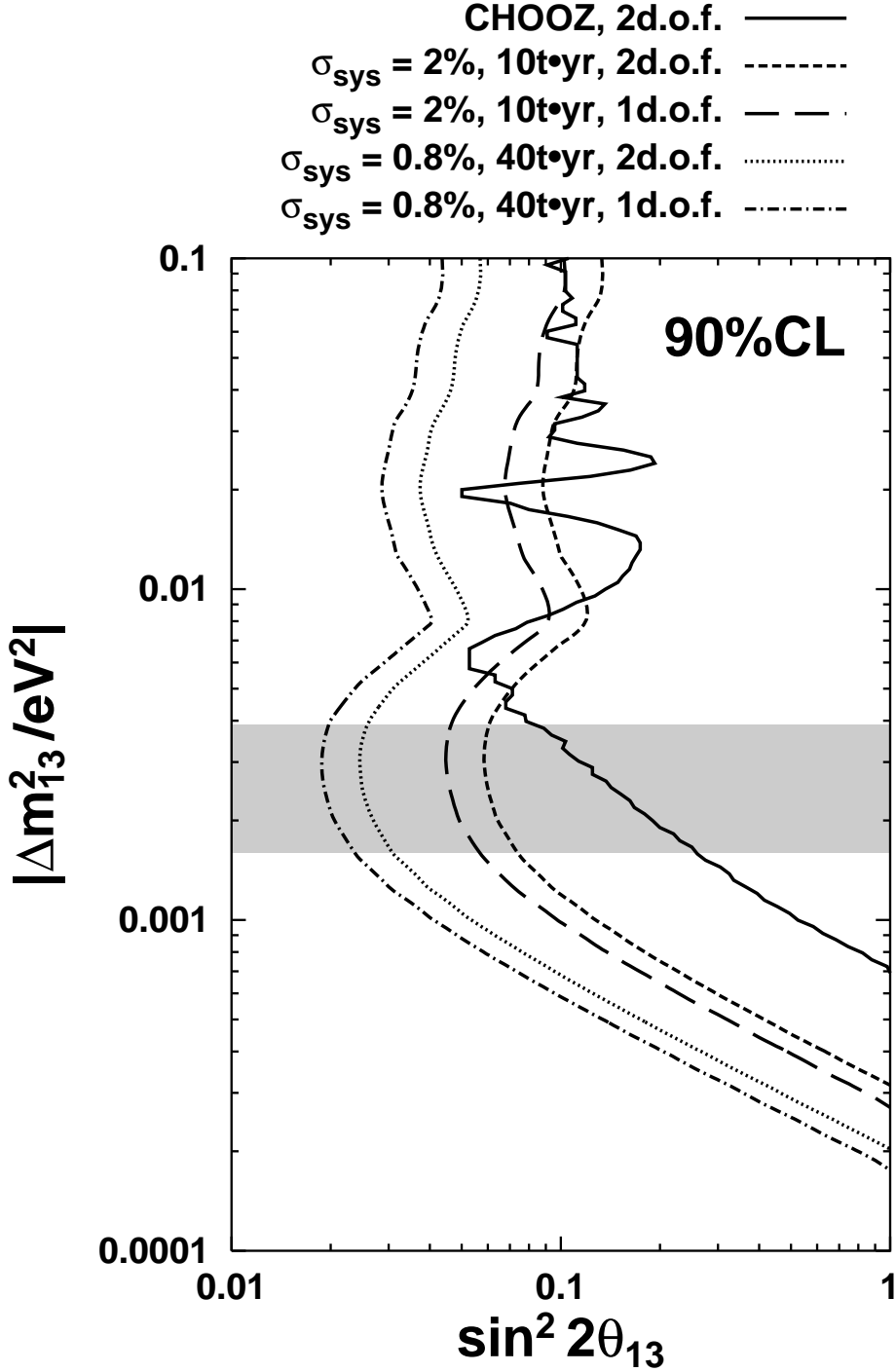}
\vglue 2.0cm
\caption{
Shown are the 90\%~CL\ exclusion limits on $\sin^2{2\theta_{13}}$
which can be placed by the reactor measurement as described in Sec.~III\@.
From the left to right,
the dash-dotted and the thin-dotted
(the long-dashed and short-dashed) lines
are based on analyses with 1 and 2 degrees of freedom (see the text),
respectively for
$\sigma_{\text{sys}}$=0.8\,\%, 40\,t$\cdot$yr
($\sigma_{\text{sys}}$=2\,\%, 10\,t$\cdot$yr).
The solid line is the CHOOZ result, and the 90\,\%\,CL interval
$1.6\times10^{-3}\,\mbox{eV}^2\le
\Delta m^2_{31}\le 3.9\times10^{-3}\,\mbox{eV}^2$
of the Super-Kamiokande atmospheric neutrino data
is shown as a shaded strip.
}
\label{exclude}
\end{figure}

First, let us calculate how much we could constrain
$\sin^2{2\theta_{13}}$.
Unlike the analysis in \cite{bugey} which
uses the ratio of the numbers of events
at the near and the far detectors,
we use the difference of the numbers of events
$N_i(L_2)-(L_1/L_2)^2N_i(L_1)$, because
the statistical analysis with ratios is complicated.
(See, e.g., \cite{Fogli:1995xu}.)
The definition of $\Delta\chi^2$, which
stands for the deviation from the best fit point
(non-oscillation point)
is given by
\begin{eqnarray}
%&&
\hspace*{-20mm}
\displaystyle
&{\ }&\Delta\chi^2(\sin^2{2\theta_{13}}, |\Delta m^2_{31}|)\nonumber\\
&\equiv&
 \sum_{i=1}^{14} \frac{\left\{
\left[
N_{i(0)}(L_2)- \left({L_1 \over L_2}\right)^2N_{i(0)}(L_1)\right]
-\left[N_i(L_2) - \left({L_1 \over L_2}\right)^2N_i(L_1)\right]
\right\}^2}
{{N_i}_{(0)}(L_2) + \left({L_1 \over L_2}\right)^4{N_i}_{(0)}(L_1)
+(\sigma_{\text{sys}}^{\text{bin}})^2{N_i}_{(0)}^2(L_2)},
\label{chi2}
\end{eqnarray}
%\\[3mm]
%&&
%\hspace*{-20mm}
\begin{eqnarray}
 N_i(L_j) \equiv N_i(\sin^2{2\theta_{13}}, |\Delta m^2_{31}|; L_j) ,\ \
 {N_i}_{(0)}(L_j) \equiv N_i(0, 0; L_j) ,
\nonumber
\end{eqnarray}
where $\sigma_{\text{sys}}^{\text{bin}}$ is
the relative systematic error for each bin
which is assumed to be the same for all bins
and $N_i(\sin^2{2\theta_{13}}, |\Delta m^2_{31}|)$ denotes
the theoretical number of $\bar{\nu}_e$ events
within the $i$th energy bin.
In principle both the systematic errors
$\sigma_{\text{abs.sys}}^{\text{bin}}$ (absolute normalization)
and $\sigma_{\text{sys}}^{\text{bin}}$ (relative normalization)
appear in the denominator of
(\ref{chi2}), but by taking the difference, we have
$(1+\sigma_{\text{abs.sys}}^{\text{bin}})[(1+\sigma_{\text{sys}}^{\text{bin}})
N_i(L_2)-(L_1/L_2)^2N_i(L_1)]
-[N_i(L_2)-(L_1/L_2)^2N_i(L_1)]=\sigma_{\text{sys}}^{\text{bin}} N_i(L_2)
+\sigma_{\text{abs.sys}}^{\text{bin}}
[N_i(L_2)-(L_1/L_2)^2N_i(L_1)]$ which indicates that
the systematic error is dominated by the relative
error $\sigma_{\text{sys}}^{\text{bin}}$, as the second term
$[N_i(L_2)-(L_1/L_2)^2N_i(L_1)]$ is supposed to be small.
In fact we have explicitly verified numerically
that the presence of
$(\sigma_{\text{abs.sys}}^{\text{bin}})^2
[N_i(L_2)-(L_1/L_2)^2N_i(L_1)]^2$ in the denominator of
(\ref{chi2}) does not affect
any result.
From the assumption that the relative systematic error
for each bin
is distributed equally into bins,
$\sigma_{\text{sys}}^{\text{bin}}$
is estimated from the relative systematic error $\sigma_{\text{sys}}$
for the total number of events by
\begin{eqnarray}
(\sigma_{\text{sys}}^{\text{bin}})^2 =
\sigma_{\text{sys}}^2
\frac{(N_{(0)}^{\text{tot}}(L_2))^2}
{\sum_i {N_i}_{(0)}^2(L_2)},
\ \ \
N_{(0)}^{\text{tot}}(L_2) \equiv \sum_i {N_i}_{(0)}(L_2),
\label{binerror}
\end{eqnarray}
since the uncertainty squared of the total number of events
is obtained by adding up the
bin-by-bin systematic errors
$(\sigma_{\text{sys}}^{\text{bin}})^2 {N_i}_{(0)}^2(L_2)$;
The ratio $\sigma_{\text{sys}}^{\text{bin}} / \sigma_{\text{sys}}$
is about 3 in our analysis.
In Fig.~\ref{exclude}, the \mbox{90\,\%\,CL} exclusion limits,
which corresponds to $\Delta \chi^2 = 2.7$ for 1 degree of freedom,
are presented for two cases:
a \mbox{10\,ton-year} measurement
with the $2\,\%$ systematic error of the total number of events
and a \mbox{40\,ton-year} measurement
with the $0.8\,\%$ error.
The figure shows that it is possible to measure
$\sin^2{2\theta_{13}}$ down to 0.02 at the maximum sensitivity
with respect to $|\Delta m^2_{31}|$,
and to 0.04 for larger $|\Delta m^2_{31}|$
by a \mbox{40\,ton-year} measurement, provided the quoted values of
the systematic errors are realized.
The CHOOZ result \cite{CHOOZ} is also depicted in Fig.~\ref{exclude}.
For a fair comparison with the CHOOZ contour, we also present
in Fig.~\ref{exclude} the results of analysis with 2 degrees of freedom,
which correspond to $\Delta \chi^2 = 4.6$ for \mbox{90\,\%\,CL},
without assuming any precise knowledges on $|\Delta m^2_{31}|$.

\begin{figure}[h]
%\vglue 4.3cm
\vglue -1.5cm
%\hglue -3.5cm
\includegraphics[scale=0.57]{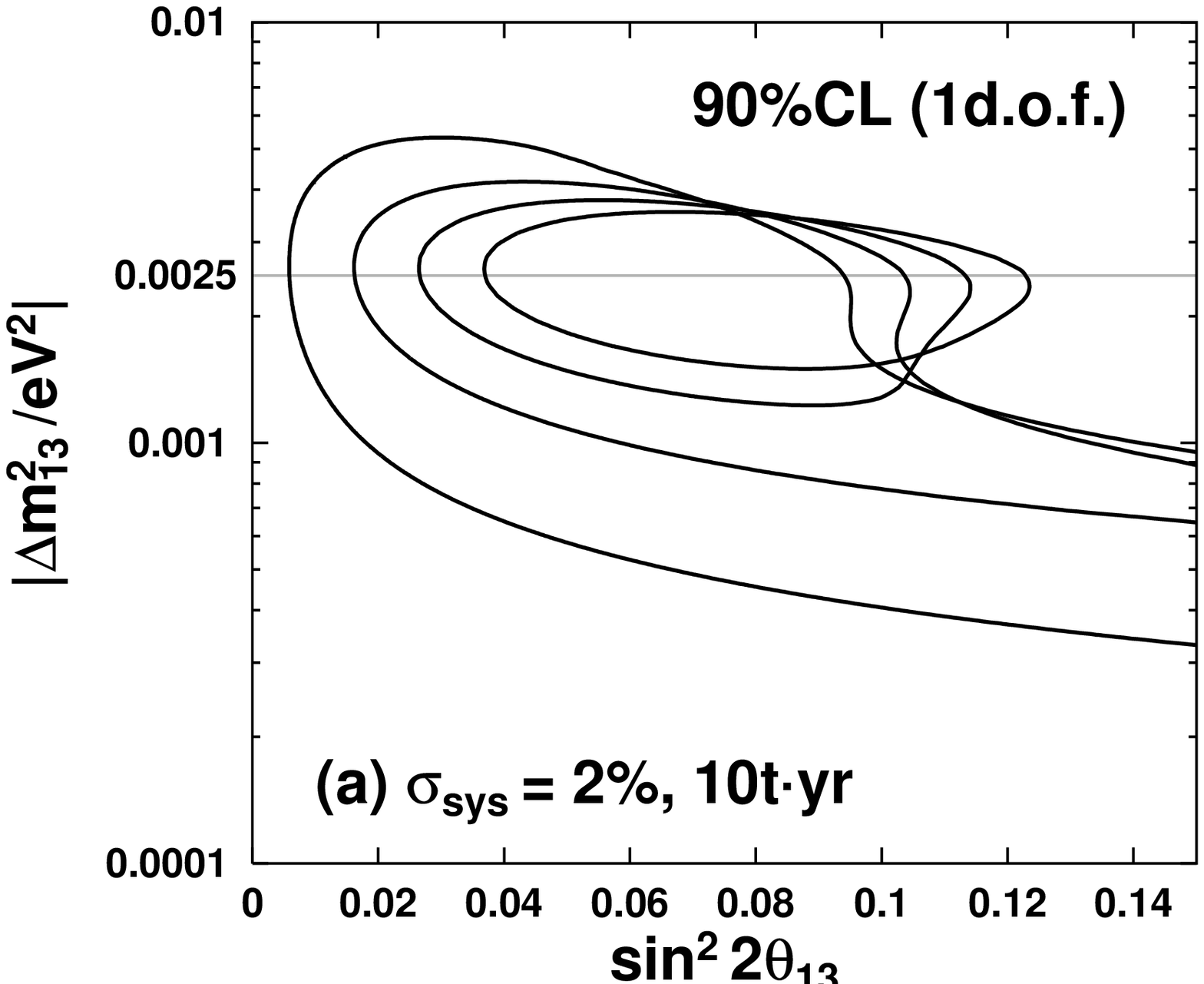}
\vglue 0.5cm
%\hglue -3.5cm
\includegraphics[scale=0.57]{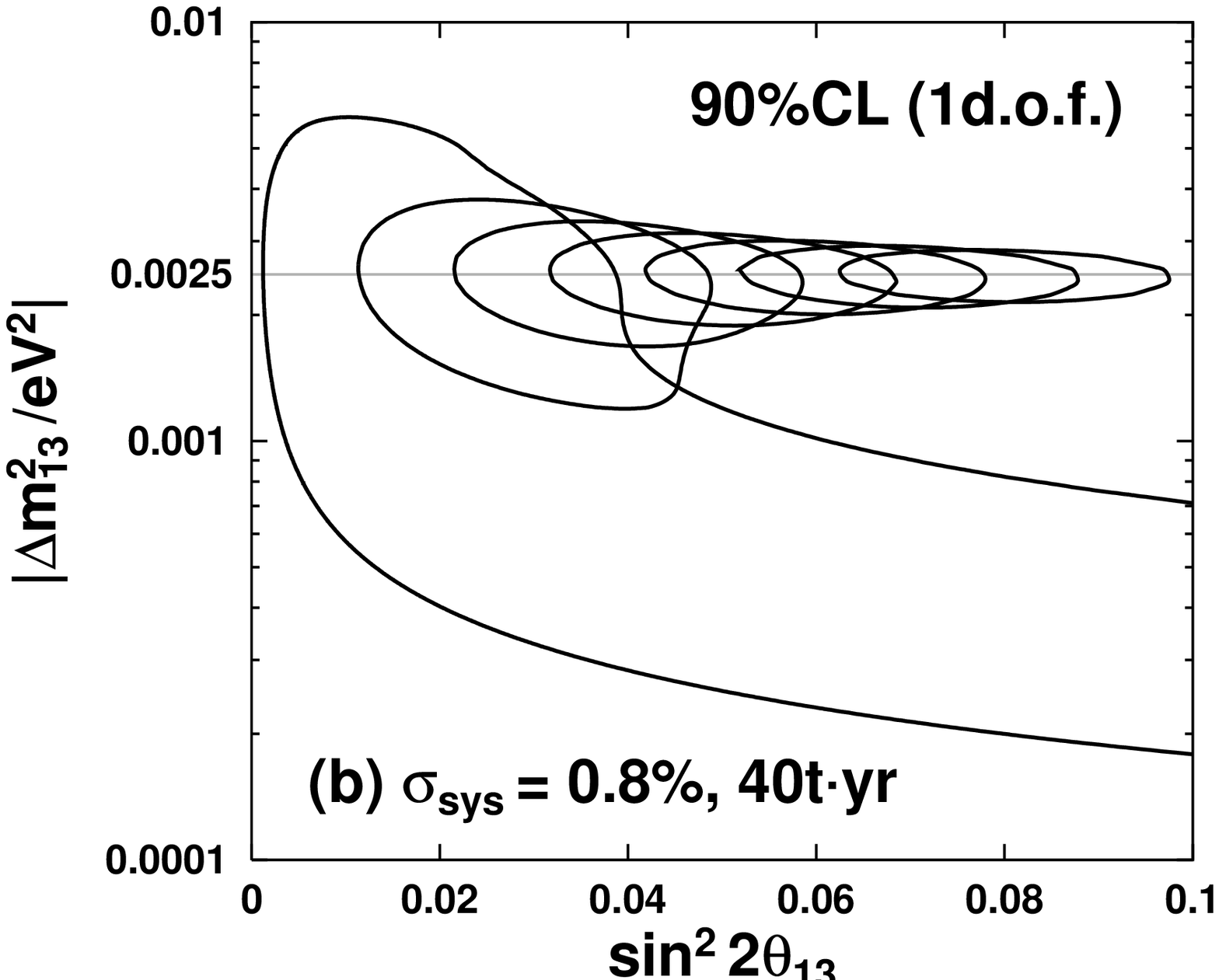}
\vglue 1.5cm
\caption{
Shown is the accuracy of determination of
$\sin^2{2\theta_{13}}$ at \mbox{90\,\%\,CL}
for the case of positive evidence
based on analysis with 1 degree of
freedom, $\Delta\chi^2 = 2.7$. Figures (a) and (b) are
for $\sigma_{\text{sys}}$=2\%, 10 t$\cdot$yr, and
$\sigma_{\text{sys}}$=0.8\%, 40 t$\cdot$yr, respectively.
The lines correspond to the best fit values of
$\sin^2{2\theta_{13}}$, from left to right,
0.05 to 0.08 in the unit of 0.01 in Fig.~\ref{allow}(a),
and
0.02 to 0.08 in the unit of 0.01 in Fig.~\ref{allow}(b).
The reference value of $|\Delta m^{2({\text{best}})}_{31}|$ is
taken to be $2.5\times10^{-3}\,{\text{eV}}^2$,
which is indicated by a gray line.
}
\label{allow}
\end{figure}

 Next, let us examine how precisely we could measure
$\sin^2{2\theta_{13}}$.
 The definition of $\Delta\chi^2$ is
\begin{eqnarray}
\hspace*{-20mm}
\displaystyle
&{\ }&\Delta\chi^2(\sin^2{2\theta_{13}}, |\Delta m^2_{31}|)\nonumber\\
&\equiv&
 \sum_{i=1}^{14} \frac{\left\{
\left[
N_{i({\text{best}})}(L_2)- \left({L_1 \over L_2}\right)^2N_{i({\text{best}})}(L_1)\right]
-\left[N_i(L_2) - \left({L_1 \over L_2}\right)^2N_i(L_1)\right]
\right\}^2}
{{N_i}_{({\text{best}})}(L_2) + \left({L_1 \over L_2}\right)^4{N_i}_{({\text{best}})}(L_1)
+(\sigma_{\text{sys}}^{\text{bin}})^2 {N_i}_{({\text{best}})}^2(L_2)} ,
\end{eqnarray}
where ${N_i}_{({\text{best}})}$ denotes $N_i$
for the set of the best fit parameters
($\sin^2{2\theta^{({\text{best}})}_{13}}$,
$|\Delta m^{2({\text{best}})}_{31}|$)
given artificially.
 $\sigma_{\text{sys}}^{\text{bin}}$ is obtained in (\ref{binerror})
by replacing ${N_i}_{(0)}$ with ${N_i}_{({\text{best}})}$
and the ratio $\sigma_{\text{sys}}^{\text{bin}} / \sigma_{\text{sys}}$
is about 3 again.
We assume that the value of $|\Delta m^2_{31}|$ is
known to a precision of $10^{-4}\,\mbox{eV}^2$ by the JHF phase I
by the time the reactor measurement is actually utilized to
solve the degeneracy. Then, we rely on the analysis with
1 degree of freedom, fixing $|\Delta m^2_{31}|$ as
$|\Delta m^{2({\text{best}})}_{31}| = 2.5\times10^{-3}\,\mbox{eV}^2$.
The \mbox{90\,\%\,CL} allowed regions of 1 degree of freedom,
whose bounds correspond to $\Delta \chi^2$ = 2.7,
are presented in Fig.~\ref{allow} for the values of
$\sin^2{2\theta^{({\text{best}})}_{13}}$ from 0.05 to 0.08
(0.02 to 0.08) in the unit of 0.01 in the case of a
\mbox{10\,ton-year} (\mbox{40\,ton-year}) measurement
with systematic error $\sigma_{\text{sys}}$ = $2.0 (0.8)\,\%$.
We can read off the errors at \mbox{90\,\%\,CL}
in $\sin^2{2\theta_{13}}$
and it is almost independent of the central
value $\sin^2{2\theta^{({\text{best}})}_{13}}$.
Thus, we have
\begin{eqnarray}
 \sin^2{2\theta_{13}} &=& \sin^2{2\theta^{({\text{best}})}_{13}}\pm0.043
\qquad(\mbox{\rm at 90\,\%\,CL,~d.o.f. = 1})
\nonumber\\
&{\ }&\mbox{\rm for}~~\sin^2{2\theta^{({\text{best}})}_{13}}\agt0.05
\nonumber
\end{eqnarray}
in the case of
$\sigma_{\text{sys}} = 2\,\%$ with a \mbox{10\,ton-year} measurement, and
\begin{eqnarray}
 \sin^2{2\theta_{13}} &=& \sin^2{2\theta^{({\text{best}})}_{13}}\pm0.018
\qquad(\mbox{\rm at 90\,\%\,CL,~d.o.f. = 1})
\nonumber\\
&{\ }&\mbox{\rm for}~~\sin^2{2\theta^{({\text{best}})}_{13}}\agt0.02
\nonumber
\end{eqnarray}
in the case of
$\sigma_{\text{sys}} = 0.8\,\%$ with a \mbox{40\,ton-year} measurement.

\section{The problem of the ($\theta_{13}$, $\delta$, $\theta_{23}$,
$\Delta m_{31}^2$) parameter degeneracy}

We explore in this and the following sections the possible
significance of reactor measurements of $\theta_{13}$ in
the context of the problem of the parameter degeneracy.
We show that the reactor measurement of $\theta_{13}$ can resolve
the degeneracy at least partly if the measurement is
sufficiently accurate.
Toward the goal we first explain what is the problem of
the parameter degeneracy in long-baseline neutrino oscillation experiments.
It is a notorious problem; a set of measurements of
the $\nu_\mu$ disappearance probability
and the appearance oscillation probabilities of
$\nu_{\mu} \rightarrow \nu_{e}$
and $\bar{\nu}_{\mu} \rightarrow \bar{\nu}_{e}$,
no matter how accurate they may be, does not allow unique determination
of $\theta_{13}$, $\delta$, and $\theta_{23}$.
The problem was first recognized in the form of intrinsic
degeneracy between the two sets of solutions of
($\theta_{23}$, $\theta_{13}$) for a given set of
measurements in two different channels
$\nu_{\mu} \rightarrow \nu_e$ and
$\nu_{\mu} \rightarrow \nu_{\tau}$ \cite{FL96}.
It was then observed independently that the similar degeneracy
of solutions of ($\theta_{13}$, $\delta$)
exists in measurement of $\nu_e$ appearance in neutrino and
antineutrino channels \cite{Burguet-Castell:2001ez}.
They made the first systematic analysis of the degeneracy
problem.
It was noticed that the degeneracy is further duplicated
provided that the two neutrino mass patterns,
the normal ($\Delta m^2_{31} > 0$) and
the inverted ($\Delta m^2_{31} < 0$) hierarchies,
are allowed \cite{Minakata:2001qm}.
Finally, it was pointed out that the degeneracy can be
maximally eight-fold \cite{Barger:2001yr}.
Analytic structure of the degenerate solutions
was worked out in a general setting in \cite{Minakata:2002qi}.

To illuminate the point, let us first restrict our
treatment to a relatively short baseline experiment such as
the CERN-Frejus project \cite{SPL}.
In this case, one can use the vacuum oscillation approximation
for the disappearance and the appearance probabilities.
From the general formula (\ref{Pgeneral}) we have
\begin{eqnarray}
1- P(\nu_{\mu} \rightarrow \nu_{\mu})
&=& 4\sum_{j<k}|U_{\mu j}|^2|U_{\mu k}|^2\sin^2\left(
{\Delta m^2_{jk}L \over 4E}\right)
\nonumber \\
&=& \sin^2{2 \theta_{23}}
\sin^2{\frac{\Delta_{31}}{2}}
\nonumber \\
&-&
\left(
\frac{1}{\,2\,} c^2_{12} \sin^2{2 \theta_{23}}
-
s_{13} s^2_{23} \sin{2 \theta_{23}}
\sin{2 \theta_{12}} \cos{\delta}
\right)
\sin{\Delta_{21}}\sin{\Delta_{31}}
\nonumber \\
&+&
O(\epsilon^2) + O(s^2_{13}),
\label{P_mumu}
\end{eqnarray}
\begin{eqnarray}
\left\{ \begin{array}{c}
P(\nu_{\mu} \rightarrow \nu_{e})\\
P(\bar{\nu}_{\mu} \rightarrow\bar{\nu}_{e})
\end{array}\right\}
&=&-4\sum_{j<k}\mbox{\rm Re}\left(U_{\mu j}U_{e j}^\ast
U_{\mu k}^\ast U_{e k}\right)\sin^2\left(
{\Delta m^2_{jk}L \over 4E}\right)\nonumber\\
&{\ }&\mp2\sum_{j<k}\mbox{\rm Im}\left(U_{\mu j}U_{e j}^\ast
U_{\mu k}^\ast U_{e k}\right)\sin\left(
{\Delta m^2_{jk}L \over 2E}\right),\nonumber \\
&=&  s^2_{23} \sin^2{2 \theta_{13}}
\sin^2{\frac{\Delta_{31}}{2}}\nonumber \\
&+&
\frac{1}{\,2\,} J_r
\sin{\Delta_{21}} \sin{\Delta_{31}}
\cos{\delta}
\nonumber \\
&\mp&
J_r
\sin{\Delta_{21}} \sin^2{\frac{\Delta_{31}}{2}}
\sin{\delta} +
O(\epsilon s^2_{13}),
\label{P_mue}
\end{eqnarray}
where $\epsilon \equiv \Delta m^2_{21}/|\Delta m^2_{31}|$,
$J_r \equiv \sin{2 \theta_{23}}\sin{2 \theta_{12}} c^2_{13}s_{13}$, and
the parametrization (\ref{MNSmatrix}) has been used in the second line
in each formula.
The minus and plus signs in front of
$\sin{\delta}$ term in (\ref{P_mue}) correspond to neutrino and
antineutrino channels, respectively.
An explicit perturbative computation in \cite{AKS}
indicates that the matter effect enters into the expression
in a particular combination with other quatities
(in the form of $s^2_{13} aL /\Delta_{31}$), so that
the effect is small.
By the disappearance measurement at JHF, for example,
$\sin^2{2 \theta_{23}}$ and $|\Delta m^2_{31}|$
will be determined with accuracies of \mbox{1\,\%} for
$0.92\le\sin^22\theta_{23}\le1.0$ (Fig.~11 in \cite{JHF}),
and \mbox{4\,\%} level, respectively \cite{JHF}.\footnote{
%%%%%%%%%%%%%% footnote %%%%%%%%%%%%%%%%
Usually one thinks of determining not $|\Delta m^2_{31}|$ but
$|\Delta m^2_{32}|$ by the disappearance measurement. But,
it does not appear possible to resolve difference between
these two quantities because one has to achieve resolution of
order $\epsilon$ for the reconstructed neutrino energy.
}
If $\theta_{23}$ is not maximal, then we have two solutions for $\theta_{23}$
($\theta_{23}$ and $\pi/2 - \theta_{23}$),
even if we ignore the uncertainty in the determination of
$\sin^2{2 \theta_{23}}$. For example,
if $\sin^2{2 \theta_{23}} = 0.95$, which is perfectly allowed by
the most recent atmospheric neutrino data \cite {shiozawa},
then $s^2_{23}$ can be either 0.39 or 0.61.
Since the dominant term in the appearance probability
depends upon $s^2_{23}$ instead of
$\sin^2{2 \theta_{23}}$, it leads to
\mbox{$\pm20$\,\%} difference in the number of appearance events
in this case.
On the other hand, in the case of maximal mixing,
it still leaves a rather wide range of $\theta_{23}$,
despite such fantastic accuracy of the measurement.
\mbox{1\,\%} accuracy in $\sin^22\theta_{23}$ implies
about \mbox{10\,\%} uncertainty in
$s^2_{23}$.
Thus, whenever we try to determine $\sin^2{2 \theta_{13}}$
from the appearance measurement,
we have to face the ambiguity due to
the two-fold nature of the solution for $s^2_{23}$.

\begin{figure}[h]
\vglue 0.6cm
\hglue -1.cm
\includegraphics[scale=0.6]{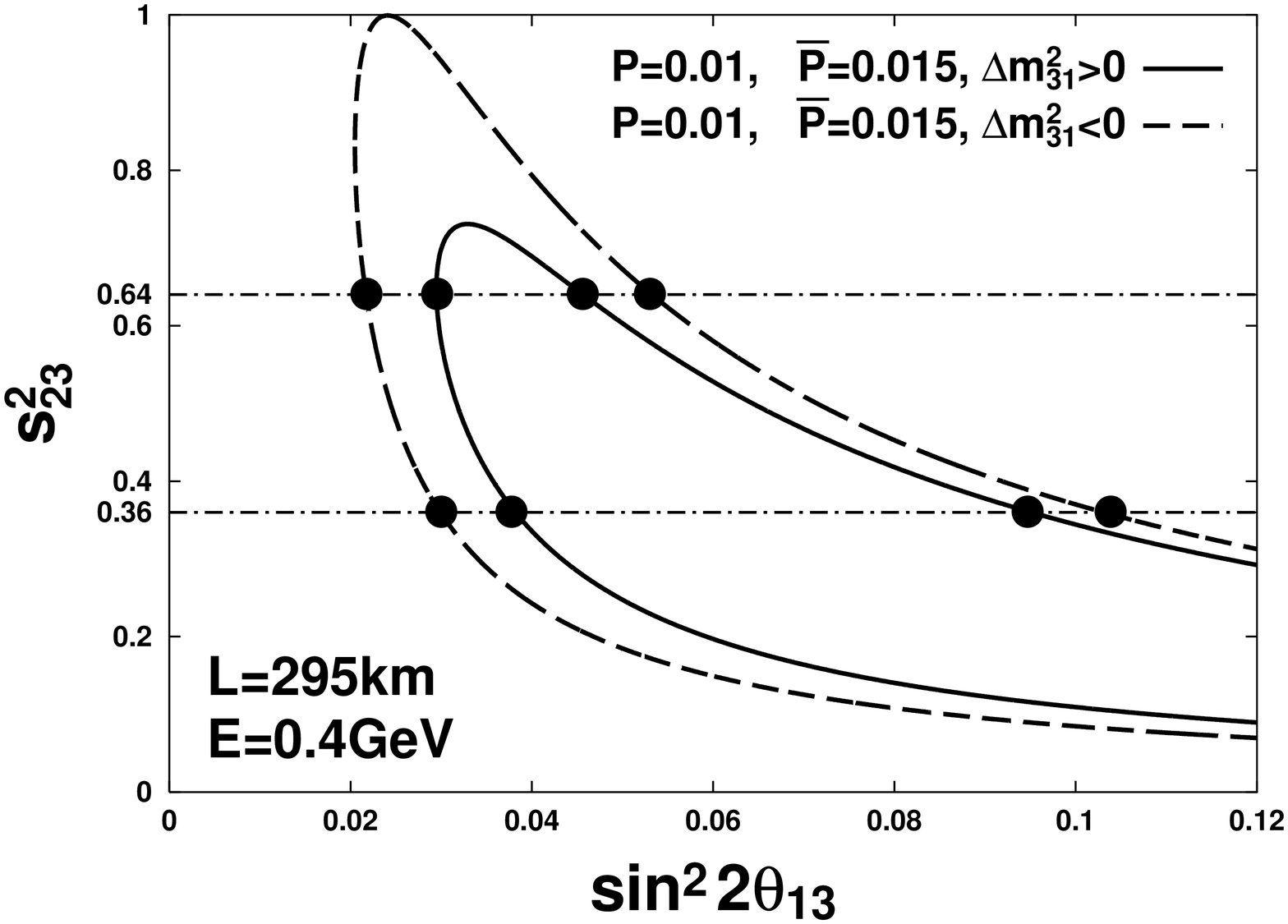}
\vglue 0.7cm
\caption{
Depicted in the $\sin^2{2\theta_{13}}$--$s^2_{23}$ plane
are the contours determined by arbitrarily given
values of the appearance probabilities
$P\equiv P(\nu_\mu\rightarrow\nu_e)=0.01$ and
$\bar{P}\equiv P(\bar{\nu}_\mu\rightarrow\bar{\nu}_e)=0.015$
with $E/L$ off the oscillation maximum
($|\Delta_{31}|\ne\pi$) at the JHF experiment.
Here, $s^2_{23} \equiv \sin^2{\theta_{23}}$. 
The solid and the dashed lines correspond to positive
and negative $\Delta m^2_{31}$, respectively.
The dash-dotted lines represent the boundary of the region
$0.36 \leq s^2_{23} \leq 0.64$ which is
presently allowed by the atmospheric neutrino data,
$0.92 \leq \sin^2{2\theta_{23}} \leq 1$.
As indicated in the figure, there are four solutions for
each $s^2_{23}$, and altogether there are eight solutions
as denoted by blobs for any values of $\theta_{23} \neq \pi/4$.
The oscillation parameters are taken as follows:
$\Delta m^2_{31}=2.5\times10^{-3}$eV$^2$,
$\Delta m^2_{21}=6.9\times10^{-5}$eV$^2$,
$\tan^2\theta_{12}=0.38$.  The Earth density
is taken to be $\rho$=2.3 g/cm$^3$.
}
\label{degen1}
\end{figure}

Let us discuss the simplest possible case, the LOW or the vacuum
(VAC) oscillation solution of the solar neutrino problem.
(See e.g., \cite{Barenboim:2002nv} for a recent discussion.)
In this case, one can safely ignore terms of order $\epsilon$
in (\ref{P_mumu}) and (\ref{P_mue}). Then we are left with
only the first terms in the right-hand-side of these equations,
the one-mass scale dominant vacuum oscillation probabilities.
Now let us define the symbols
$x=\sin^2{2 \theta_{13}}$ and $y=s^2_{23}$.
Then, (\ref{P_mumu}) and (\ref{P_mue}) take the forms
$y=y_1$ or $y_2$ (corresponding to two solutions of $s^2_{23}$)
and $xy=constant$, respectively, for given values of the
probabilities. It is then obvious that there are
two crossing points of these curves.
This is the simplest version of the ($\theta_{13}, \theta_{23}$)
degeneracy problem.
%
%The feature as well as more elaborate ones, on which we discuss
%immediately below, are pictorially displayed in Fig.~\ref{degen1}.
We next discuss what happens if $\epsilon$ is not
negligible though small: the case of LMA solar neutrino solution.
In this case, the appearance curve, $xy=constant$, split into
two curves (though they are in fact connected at their maximum
value of $s^2_{23}$) because of the two degenerate solution
of the set ($\delta$, $\theta_{13}$)
that is allowed for a given set of values of $s_{23}^2$,
$P(\nu_{\mu} \rightarrow \nu_{e})$ and
$P(\bar{\nu}_{\mu} \rightarrow \bar{\nu}_{e})$.
Then, we have, in general, four crossing points on the \mbox{x--y} plane
for a given value of
$\sin^2{2 \theta_{23}}$, the four-fold degeneracy.
Simultaneously, the two $y= constant$ lines are slightly
tilted and the splitting between two curves becomes larger
at larger $\sin^2{2 \theta_{13}}$, though the effect is
too tiny to be clearly seen.
If the baseline distance is longer, the Earth matter effect
comes in and further splits each appearance contour
into two, depending upon the sign of $\Delta m^2_{31}$.
Then, we have four curves
(or, two continuous contours each of which intersects with
$y = constant$ line twice) and hence there are
eight solutions as displayed in Fig.~\ref{degen1}.\footnote
%%%%%%%%%%%%%%%%   Footnote   %%%%%%%%%%%%%%%%%%%%
{
The readers might be curious about the feature that the two
contours are connected with each other at a large $s^2_{23}$
point. Because $\delta$ is a phase variable,
the contours must be closed as $\delta$ varies.
}
This is a simple pictorial representation of the maximal
eight-fold parameter degeneracy \cite {Barger:2001yr}.
To draw Fig.~\ref{degen1}, we have calculated disappearance and
appearance contours by using the approximate formula derived by
Cervera {\it et al.} \cite{golden}. We take the baseline distance
and neutrino energy as \mbox{$L=295$\,km} and \mbox{$E=400$\,MeV}
with possible
relevance to JHF project \cite{JHF}. The Earth matter density is
taken to be $\rho = 2.3\,\mbox{g}\cdot\mbox{cm}^{-3}$ based on
the estimate given in \cite{KSmpl99}. The electron fraction
$Y_e$ is taken to be 0.5.
We assume, for definiteness, that a long-baseline
disappearance measurement has resulted in
$\sin^2{2 \theta_{23}} = 0.92$ and
$\Delta m^2_{31} = 2.5 \times 10^{-3}\,\mbox{eV}^2$.
For the LMA solar neutrino parameters we take
$\tan^2\theta_{12}=0.38$ and
$\Delta m^2_{21}=6.9\times10^{-5}\,\mbox{eV}^2$
\cite{Fukuda:2002pe}.
We take the values of these parameters and the matter density
throughout this paper unless otherwise stated.
The qualitative features of the figure remain unchanged even if
we employ values of the parameters obtained by other analyses.

\section{Resolving the parameter degeneracy by reactor measurement
of $\theta_{13}$}

Now we discuss how reactor experiments can contribute to
resolve the parameter degeneracy. To make our discussion as
concrete as possible we use the particular long-baseline experiment,
the JHF experiment \cite{JHF}, to illuminate the complementary
role played by reactor and long-baseline experiments.
It is likely that the experiment will be carried out at around
the first oscillation maximum ($|\Delta_{31}| = \pi$)
for a number of reasons:
the dip in energy spectrum in disappearance channel is the deepest,
the number of appearance events are nearly maximal \cite {JHF},
and the two-fold degeneracy in $\delta$ becomes simple
($\delta \leftrightarrow \pi - \delta$) for each mass hierarchy
\cite{KMN02,Barger:2001yr}.\footnote{
%%%%%%%%%%%%%%%%% footnote %%%%%%%%%%%%%%%%%%%%
In order to have this reduction, one has to actually tune the
energy spectrum so that
$\cos{\delta}$ term in (\ref{P_mue}) averaged over the energy
with the neutrino flux times the cross section vanishes, which
is shown to be doable in \cite{KMN02}.
}
With the distance \mbox{$L=295$\,km}, the oscillation maximum is
at around \mbox{$E=600$\,MeV}\@. We take the same mixing parameters as
those used in Fig.~\ref{degen1}.

\subsection{Illustration of how reactor measurement helps resolve the
($\theta_{13}$, $\theta_{23}$) degeneracy}

\begin{figure}[h]
\vglue 0.6cm
\hglue -1.cm
\includegraphics[scale=0.6]{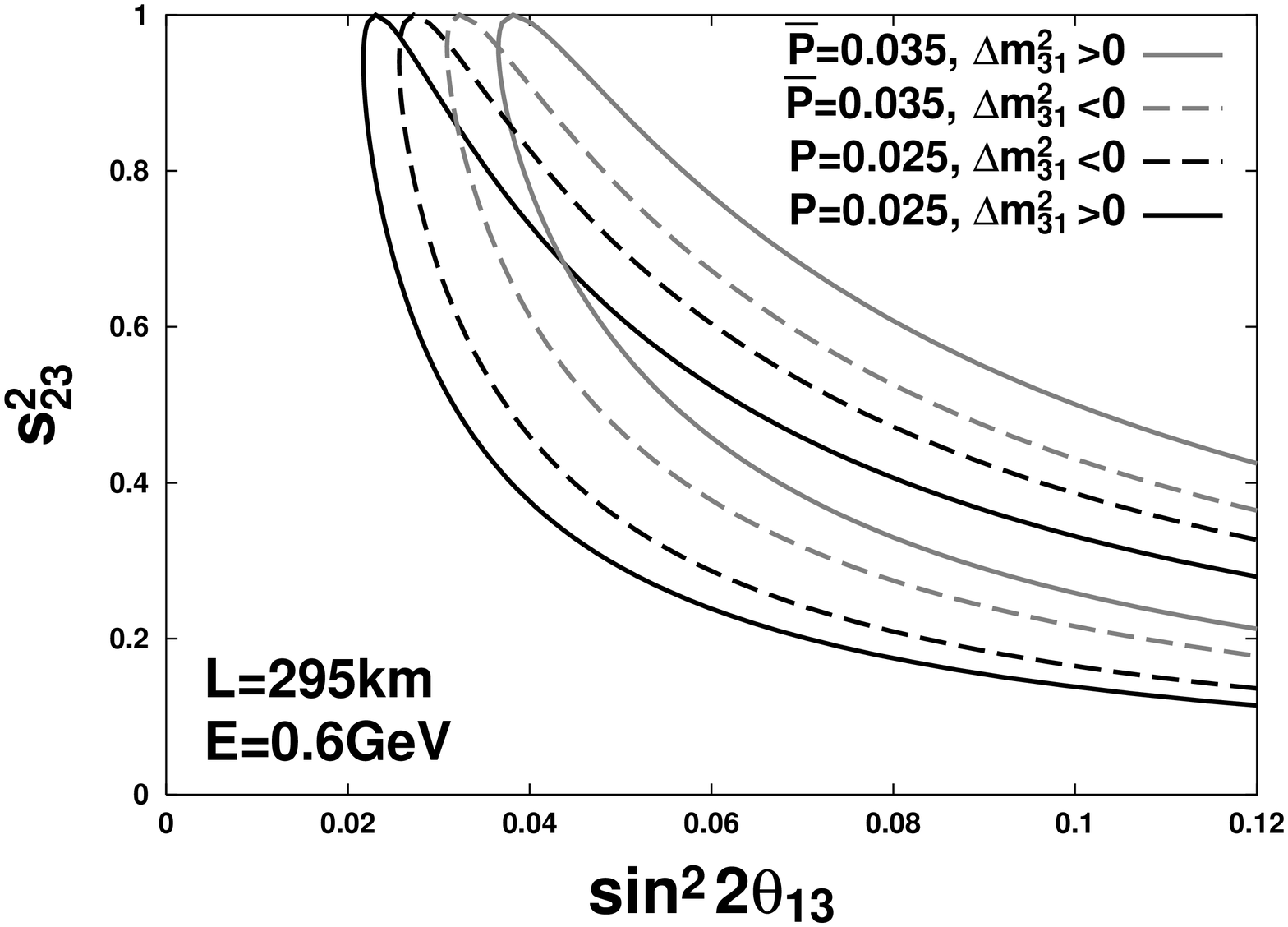}
\vglue 0.7cm
\caption{
The allowed regions are shown in the $\sin^2{2\theta_{13}}$--$s^2_{23}$
plane determined with a given value of
$P \equiv P(\nu_\mu\rightarrow\nu_e)$ alone (in this case $P=0.025$),
or $\bar{P} \equiv P(\bar{\nu}_\mu\rightarrow\bar{\nu}_e)$ alone
(in this case $\bar{P}=0.035$)
at the oscillation maximum $|\Delta_{31}|=\pi$ of the JHF experiment.
Each allowed region is the area bounded by
the black solid
(for $\Delta m^2_{31}>0$ with $P$ only),
the black dashed
(for $\Delta m^2_{31}<0$ with $P$ only),
the gray solid
(for $\Delta m^2_{31}>0$ with $\bar{P}$ only),
the gray dashed
(for $\Delta m^2_{31}<0$ with $\bar{P}$ only),
respectively, where the line with a definite value of the CP
phase $\delta$ sweeps out each region
as $\delta$ varies from $0$ to $2\pi$.
The oscillation parameters and the
Earth density are the same as those in Fig.~\ref{degen1}.
}
\label{degen2}
\end{figure}

\begin{figure}[h]
\vglue -1.5cm
\hglue -1.cm
\includegraphics[scale=0.6]{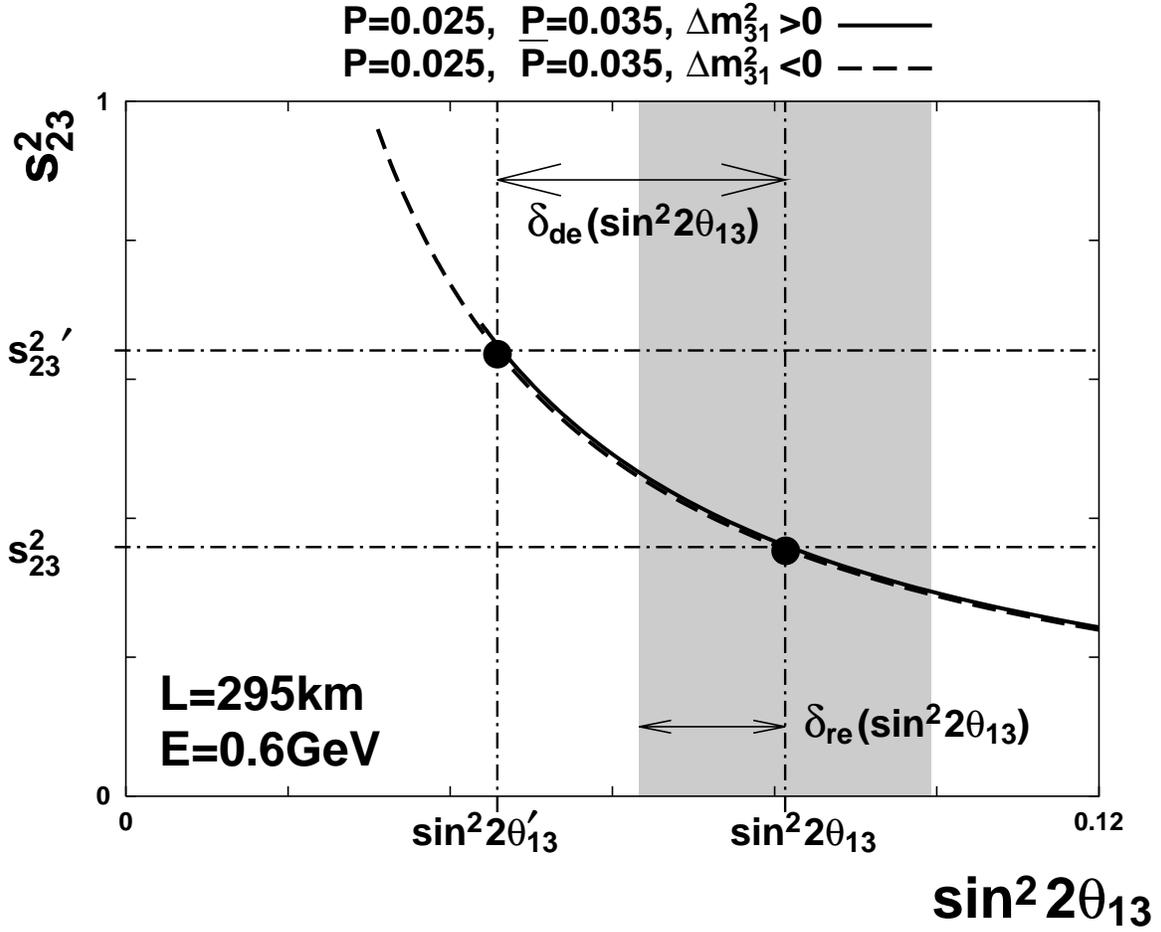}
%\vglue -0.4cm
\caption{
The allowed region in the $\sin^2{2\theta_{13}}$--$s^2_{23}$
plane becomes a line when both $P(\nu_\mu\rightarrow\nu_e)$ and
$P(\bar{\nu}_\mu\rightarrow\bar{\nu}_e)$ are given
(in this case $P(\nu_\mu\rightarrow\nu_e)=0.025$,
$P(\bar{\nu}_\mu\rightarrow\bar{\nu}_e)=0.035$)
at the oscillation maximum
($|\Delta_{31}|=\pi$, $E=0.6$ GeV for the JHF experiment),
as indicated in the figure.
The solid and the dashed lines are for $\Delta m^2_{31}>0$ and
$\Delta m^2_{31}<0$ cases, respectively.
Assuming $\theta_{23} \neq \pi/4$, two solutions of
($\sin^22\theta_{13}$, $s^2_{23}$) are plotted;
In this figure $\sin^22\theta_{23}$ is taken as 0.92.
It is assumed arbitrarily that the solution of
$\theta_{23}$ in the first octant ($\theta_{23} < \pi/4$)
is the genuine one, while the one in the second octant
($\theta_{23} > \pi/4$) with primes is the fake one.
Superimposed in the figure as a shaded region is
the anticipated error in the reactor measurement of
$\theta_{13}$ estimated in Sec.~III\@.
If the error $\delta_{\text{re}}(\sin^22\theta_{13})$
is smaller than the difference
$\delta_{\text{de}}(\sin^22\theta_{13})\equiv
|\sin^22\theta'_{13}-\sin^22\theta_{13}|$
due to the degeneracy, then the reactor experiment may
be able to resolve it.
}
\label{degen3}
\end{figure}

Let us first give an illustrative example showing
how reactor experiments could help resolve the
($\theta_{13}$, $\theta_{23}$) degeneracy.
To present a clear step-by-step explanation of the relationship
between LBL and reactor experiments, we first plot in
Fig.~\ref{degen2} the allowed regions in the
$\sin^22\theta_{13}$--$s^2_{23}$ plane
by measurements of
$P(\nu_{\mu} \rightarrow \nu_{e})$ alone
and
$P(\bar{\nu}_{\mu} \rightarrow \bar{\nu}_{e})$ alone separately.
The former is indicated by the regions bounded by black lines
and the latter by gray lines.
The solid and dashed lines are used for cases with
positive and negative $\Delta m^2_{31}$.
The values of disappearance and appearance
probabilities are chosen arbitrarily for illustrative purpose
and are given in the caption of Fig.~\ref{degen2}.
Notice that the negative $\Delta m^2_{31}$ curve is located
right (left) to the positive $\Delta m^2_{31}$ curve
in neutrino (antineutrino) channel.
A plot with only measurement in neutrino mode goes beyond
academic interest because the JHF
experiment is expected to run only with the neutrino mode
in its first phase.
We observe that there is large intrinsic uncertainty in the
$\theta_{13}$ determination due to unknown $\delta$,
the problem addressed in \cite{KMN02}.
The two regions corresponding to positive and negative
$\Delta m^2_{31}$ heavily overlap due to small matter effect.
When two measurements of $\nu$ and $\bar{\nu}$ channels
are combined, the allowed solution becomes a line
which lies inside
of the overlap of the $\nu$ and $\bar{\nu}$ regions
for each sign of $\Delta m^2_{31}$ in Fig.~\ref{degen2}.
\footnote{
In the absence of the matter effect, the reason why
the closed curve shrinks into a line at the the oscillation maximum
can be seen as follows:
By eliminating $\delta$
in (\ref{P_mue}), it is easy to show
that there are two solutions of $\sin2\theta_{13}>0$
for given values of $P$, $\bar{P}$ and $\theta_{23}$ off the
oscillation maximum ($\Delta_{31}\ne\pi$), whereas
there is only one solution of $\sin2\theta_{13}>0$
at the oscillation maximum ($\Delta_{31}=\pi$).
Even if we switch on the matter effect,
one can easily
show by using
the approximate formula in \cite{golden} that the same argument holds.
}
In Fig.~\ref{degen3} we have plotted such solutions
as two lines, one for positive $\Delta m^2_{31}$
(the solid curve) and the other for negative $\Delta m^2_{31}$
(the dashed curve) at the first oscillation maximum
$|\Delta_{31}| = \pi$.
It may appear curious that the two curves with positive and
negative $\Delta m^2_{31}$ almost overlap with each other in
Fig.~\ref{degen3}.
In fact, a slight splitting between the solid
($\Delta m^2_{31} > 0$) and dashed
($\Delta m^2_{31} < 0$) lines
is due to the fact that both $\epsilon$
and the matter effect
in the case of the JHF experiment
are small.
Thus, the degeneracy in the set ($\theta_{13}$, $\theta_{23}$)
is effectively two-fold in this case.

To have a feeling on whether the reactor experiment described
in Sec.~III will be able to resolve the degeneracy, we plot
in Fig.~\ref{degen3} two sets of degenerate solutions
by taking a particular value of $\theta_{23}$,
$\sin^2{2 \theta_{23}} = 0.92$, the lower end of the region
allowed by Super-Kamiokande.
We denote the true and fake solutions as
$(\sin^22\theta_{13}, s_{23}^2)$ and
$(\sin^22\theta_{13}^\prime, {s_{23}^2}')$, respectively,
assuming the true $\theta_{23}$ satisfies $\theta_{23}<\pi/4$.
We overlay in Fig.~\ref{degen3} a shadowed region to indicate
the accuracy to be achieved by the reactor measurement of
$\theta_{13}$.
If the experimental error $\delta_{\text{re}}(\sin^22\theta_{13})$
in the reactor measurement of $\sin^22\theta_{13}$
is smaller than the difference
\begin{eqnarray}
\delta_{\text{de}}(\sin^22\theta_{13}) \equiv
|\sin^22\theta'_{13}-\sin^22\theta_{13}|
\end{eqnarray}
due to the ($\theta_{13}$, $\theta_{23}$) degeneracy,
then the reactor experiment may resolve the degeneracy.
Notice that once the $\theta_{23}$ degeneracy is lifted
one can easily obtain four allowed sets of
($\delta$, $\Delta m_{31}^2$)
(though they are still degenerate at almost the same point
on the $\sin^22\theta_{13}$--$s_{23}^2$ plane) because the relationship
between them is given analytically in a completely general setting
\cite{Minakata:2002qi}.

\subsection{Resolving power of the ($\theta_{13}$, $\theta_{23}$)
degeneracy by a reactor measurement}

Let us make a semi-quantitative estimate of how powerful the
reactor method is for resolving the ($\theta_{13}$, $\theta_{23}$)
degeneracy.\footnote{The possibility of resolving the
($\theta_{13}$, $\theta_{23}$) by a reactor experiment was
qualitatively mentioned in \cite{FL96,Barenboim:2002nv}.}\,\footnote{
%%%%%%%%%%%%%%%%%%%%%%%% footnote %%%%%%%%%%%%%%%%%%%%%%%
An alternative way to resolve the ambiguity is to look at
$\nu_e\rightarrow\nu_\tau$ channel
%in either superbeam experiments or in neutrino factories,
because the main oscillation
term in the probability $P(\nu_e\rightarrow\nu_\tau)$
depends upon $c^2_{13}$.
Unfortunately, this idea does not appear to be explored in detail
while it is briefly mentioned in \cite{Barger:2001yr,DMM02}.}
%%%%%%%%%%%%%%%%%%%%
For this purpose, we compare in this section
the difference of the two $\theta_{13}$ solutions due to
the degeneracy with the resolving power of the reactor experiment.
We consider, for simplicity, the special case $|\Delta_{31}|=\pi$,
i.e., energy tuned at the first oscillation maximum.
%since the general case is complicated to work out.
The simplest case seems to be indicative of features of
more generic cases.

\begin{figure}[h]
\vglue 2.cm\hglue -8.8cm
\includegraphics[scale=0.6]{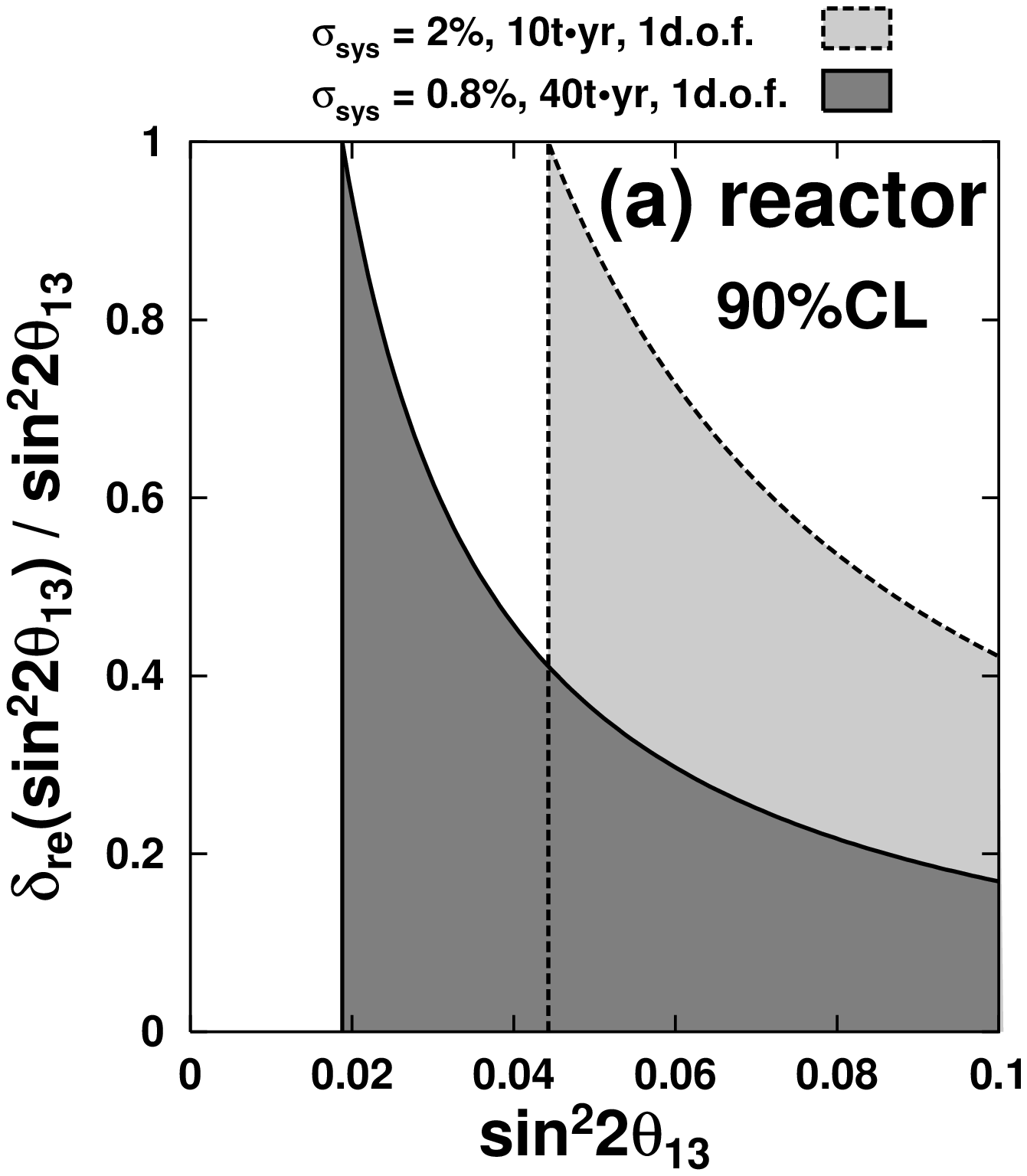}
\vglue -8.6cm\hglue 8.4cm
\includegraphics[scale=0.6]{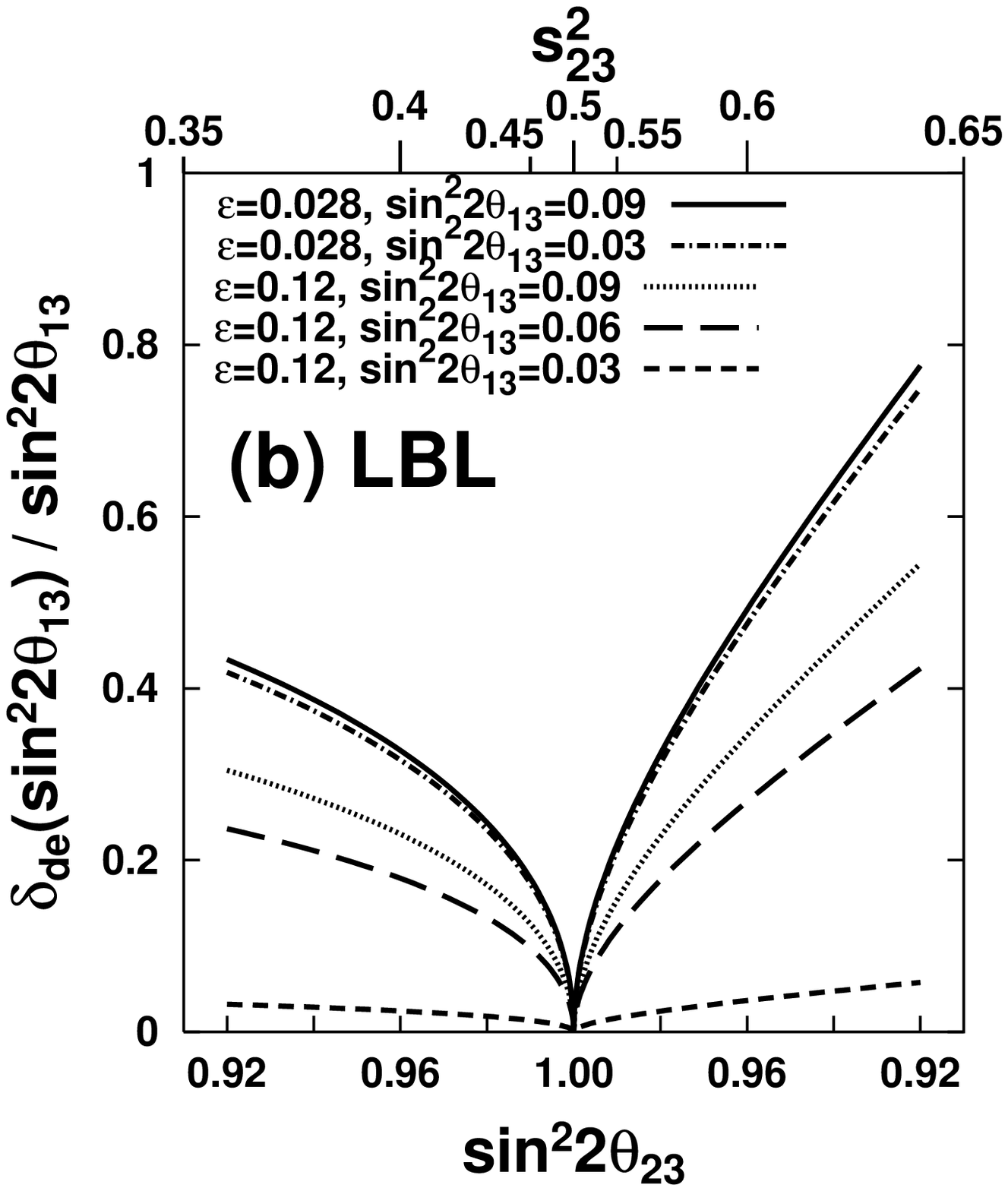}
\vglue 0.0cm
\caption{
(a) The normalized error at \mbox{90\,\%\,CL} in the reactor
measurement of $\theta_{13}$ is given for $\sigma_{\text{sys}}$=\mbox{2\,\%},
\mbox{10\,t$\cdot$yr}
(d.o.f.=1, $\delta_{\text{re}}(\sin^22\theta_{13})=0.043$)
and for $\sigma_{\text{sys}}$=\mbox{0.8\,\%}, \mbox{40\,t$\cdot$yr}
(d.o.f.=1, $\delta_{\text{re}}(\sin^22\theta_{13})=0.018$), respectively.
Notice that the degrees of freedom becomes 1
once the value of $|\Delta m^2_{31}|$ is known from JHF.\\
(b) The fractional difference
$\delta_{\text{de}}(\sin^22\theta_{13})
/\sin^22\theta_{13}$ due to the degeneracy is plotted
as a function of $\sin^22\theta_{23}$.
Here,
$\delta_{\text{de}}(\sin^22\theta_{13})\equiv
|\sin^22\theta'_{13}-\sin^22\theta_{13}|$
stands for the difference between the true solution
$\sin^22\theta_{13}$ and the fake one $\sin^22\theta'_{13}$,
and
$\epsilon\equiv\Delta m^2_{21}/|\Delta m^2_{31}|$;
$\epsilon=6.9\times10^{-5}\,\mbox{eV}^2/2.5\times10^{-3}\,\mbox{eV}^2=0.028$
is for the best fit and an extreme case with
$\epsilon=1.9\times10^{-4}\,\mbox{eV}^2/1.6\times10^{-3}\,\mbox{eV}^2=0.12$,
which is allowed at \mbox{90\,\%\,CL} (atmospheric) or \mbox{95\,\%\,CL}
(solar), is also shown for illustration.
The horizontal axis is suitably defined so that
it is linear in $\sin^22\theta_{23}$, where the left half
is for $\theta_{23}<\pi/4$ whereas the right half
is for $\theta_{23}>\pi/4$.
The solar mixing angle is taken as
$\tan^2\theta_{12}=0.38$.
$\sin^22\theta_{23}\ge0.92$ has to be satisfied
due to the constraint from the Super-Kamiokande
atmospheric neutrino data.
If the value of
$\cos^22\theta_{23}$ is large enough, the value of
$\delta_{\text{de}}(\sin^22\theta_{13})/\sin^22\theta_{13}$
increases and lies outside of the normalized error of the reactor
experiment, then the reactor result may resolve the $\theta_{23}$ ambiguity.
}
\label{delth13}
\end{figure}

\begin{figure}[h]
\hglue -1.8cm
\includegraphics[scale=0.7]{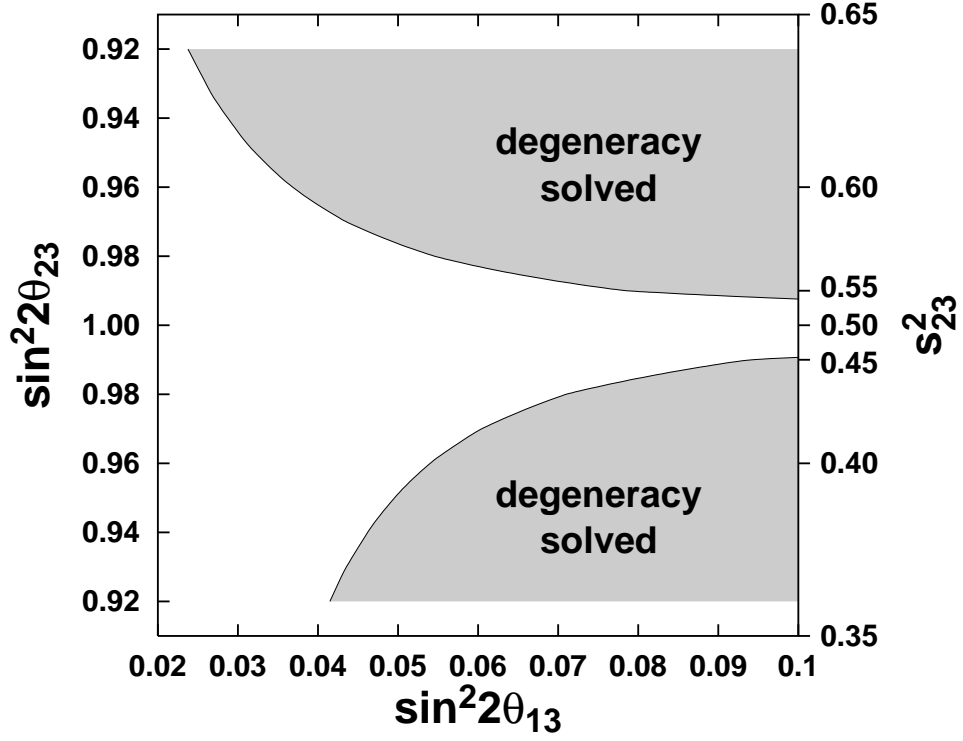}
\vglue 0.5cm
\caption{
The shadowed area stands for
the region in which $\delta_{\text{re}}(\sin^22\theta_{13})
<\delta_{\text{de}}(\sin^22\theta_{13})$
is satisfied for $\sigma_{\text{sys}}$=\mbox{0.8\,\%}, \mbox{40\,t$\cdot$yr},
d.o.f.=1 and for the best fit values of the solar and
atmospheric oscillation parameters.
In this shadowed region, the ($\theta_{13}$, $\theta_{23}$) degeneracy
may be solved.  The vertical axis is the same
as the horizontal axis of Fig. \ref{delth13}(b).
}
\label{del13tan}
\end{figure}

As we saw in the previous section, there are two solutions
of $\theta_{13}$ due to doubling of $\theta_{23}$
for a given $\sin^2{2 \theta_{23}}$ in each sign of
$\Delta m^2_{31}$.
Then, we define the fractional difference due to the degeneracy
\begin{equation}
\frac{\delta_{\text{de}}(\sin^22\theta_{13})}{\sin^22\theta_{13}} .
\label{eqn:deltade}
\end{equation}
It is to be compared with
$\delta_{\text{re}}(\sin^22\theta_{13})
/\sin^22\theta_{13}$
of the reactor experiment, where
$\delta_{\text{re}}(\sin^22\theta_{13})$
denotes the experimental uncertainty
estimated in Sec.~III\@, i.e., 0.043 or 0.018.
In Fig.~\ref{delth13}(a) we plot the normalized error
$\delta_{\text{re}}(\sin^22\theta_{13})/\sin^22\theta_{13}$
which is expected to be achieved in the reactor experiment
described in Sec.~III\@. We restrict ourselves to
the analysis with 1 degree of freedom,
because we expect that the JHF phase I will provide us accurate
information on $\Delta m^2_{31}$ by the time
when the issue is really focused on the degeneracy
in the JHF phase II\@.
The fractional difference (\ref{eqn:deltade})
can be computed from the relation \cite{Barger:2001yr}
\begin{eqnarray}
\sin^22\theta_{13}^{\prime} &=&
\sin^22\theta_{13} \tan^2\theta_{23}
+ \left({\Delta m^2_{21} \over \Delta m^2_{31}} \right)^2
{\tan^2\left(aL/2\right) \over \left(aL/\pi\right)^2}
\nonumber\\
&\times&\left[1-\left(aL/\pi\right)^2 \right]
\sin^22\theta_{12} \left(1-\tan^2\theta_{23}\right),
%\label{eqn:
\label{eqn:th23}
\end{eqnarray}
and the result for
$\delta_{\text{de}}(\sin^22\theta_{13})
/\sin^22\theta_{13}$
is plotted in Fig.~\ref{delth13}(b)
as a function of $\sin^2\theta_{23}$ for two typical values
of $\epsilon$.
We notice that the fractional differences differ by
up to a factor of \mbox{$\sim$ 2} in small $\sin^22\theta_{23}$
region between the first ($\theta_{23}<\pi/4$) and
the second octant ($\theta_{23}>\pi/4$).
For the best fit value of the two mass squared differences
$\Delta m^2_{21}$ ($6.9 \times10^{-5}\,\mbox{eV}^2$)
and $|\Delta m^2_{31}|$ ($2.5 \times10^{-3}\,\mbox{eV}^2$),
for which
$\epsilon\equiv\Delta m^2_{21}/|\Delta m^2_{31}|=0.028$,
there is little difference between the case with
$\sin^22\theta_{13}=0.03$ and the one with $\sin^22\theta_{13}=0.09$.
In this case they are all approximated by the first term in (\ref{eqn:th23})
and $\delta_{\text{de}}(\sin^22\theta_{13})
/\sin^22\theta_{13}$ depends approximately
only on $\theta_{23}$, making the analysis easier.
On the other hand, if the ratio
$\epsilon\equiv\Delta m^2_{21}/|\Delta m^2_{31}|$
is much larger than that at the best fit point, then
the second term in (\ref{eqn:th23}) is not negligible.
In Fig.~\ref{delth13}(b),
$\delta_{\text{de}}(\sin^22\theta_{13})
/\sin^22\theta_{13}$
is plotted in an extreme case of
$\epsilon=1.9\times10^{-4}$\,eV$^2/1.6\times10^{-3}$\,eV$^2=0.12$,
which is allowed at 90\% CL (atmospheric) or 95\% CL (solar),
with $\sin^22\theta_{13}=0.03, 0.06, 0.09$.  From this, we observe that
the suppression in the first term in (\ref{eqn:th23})
is compensated by the second term for $\sin^22\theta_{13}=0.03$,
i.e., the degeneracy is small and therefore
resolving the degeneracy is difficult in this case.
To clearly illustrate the resolving power of the degeneracy by
the reactor measurement, assuming the best fit value
$\epsilon=0.028$,
we plot in Fig.~\ref{del13tan}
the region where the degeneracy can be lifted in the
$\sin^22\theta_{13}$--$\sin^22\theta_{23}$ plane.
It is evident that the reactor measurement will be able to
resolve the ($\theta_{13}$, $\theta_{23})$ degeneracy
in a wide range inside its sensitivity region,
in particular for $\theta_{23}$ in the second octant.

Quantitative estimation of the significance of the fake solution
requires detailed analysis of accelerator experiments which
includes the statistical and systematic errors as well as
the correlations of errors and the parameter degeneracies,
and it will be worked out in future communication.

\section{More about reactor vs. long-baseline experiments}

The foregoing discussions in the previous section implicitly assume
that the sensitivities of reactor and LBL experiments with both
$\nu$ and $\bar{\nu}$ channels are good enough to detect effects
of nonzero $\theta_{13}$.
However, it need not be true, in particular, in coming 10 years.
To further illuminate complementary roles played by reactor and LBL
experiments, we examine their possible mutual relationship
including the cases where there is a signal in the former but
none in the latter experiments, or vice versa.
For ease of understanding by the readers,
we restrict our presentation in this section to a very intuitive
level by using a figure. It is, of course, possible to make it more
precise by deriving inequalities based on the analytic approximate
formulae \cite {golden}.
Throughout this section LBL experiments at the oscillation maximum and
$\theta_{23}=\pi/4$ are assumed.

\begin{figure}[h]
\hglue 2.0cm
\includegraphics[scale=0.8]{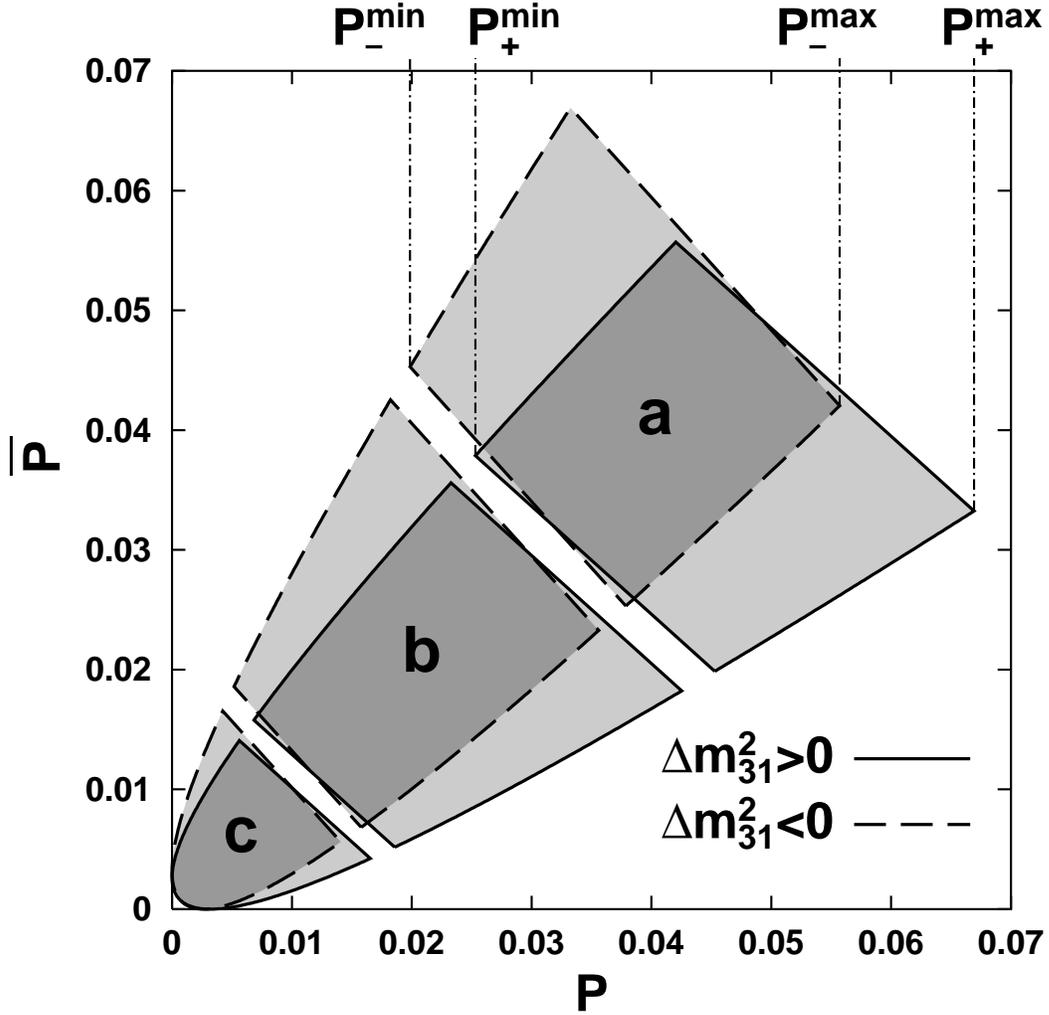}
\vglue 1.8cm
\caption
{
Predicted allowed regions are depicted in the $P$--$\bar{P}$ plane
for the JHF experiment at the oscillation maximum
after an affirmative (a negative) result of the reactor experiment
is obtained, where $P\equiv P(\nu_\mu\rightarrow\nu_e)$ and
$\bar{P}\equiv P(\bar{\nu}_\mu\rightarrow\bar{\nu}_e)$ are
the appearance probabilities, and
$\theta_{23}=\pi/4$ is assumed.  The cases a, b, c correspond to
$\sin^22\theta_{13}=0.08\pm0.018$,
$\sin^22\theta_{13}=0.04\pm0.018$,
$\sin^22\theta_{13}<0.019$, respectively.
The regions bounded by the solid lines and the
dashed lines are for
the normal hierarchy ($\Delta m^2_{31}>0$)
and the inverted hierarchy ($\Delta m^2_{31}<0$),
respectively.
Each region predicts the maximum ($P_{\pm}^{\text{max}}$)
and the minimum ($P_{\pm}^{\text{min}}$)
values of $P$
for each hierarchy ($+$ for the normal and
$-$ for the inverted hierarchy),
although $P_{\pm}^{\text{min}}$ of the region c are zero.
}
\label{envelope}
\end{figure}

If a reactor experiment sees an affirmative evidence for the disappearance
in $\bar{\nu}_e \rightarrow \bar{\nu}_e$ (the case of Reactor Affirmative),
it would be possible
to determine $\theta_{13}$ up to certain experimental errors.
In this case, the appearance probability in LBL experiment
must fall into the region
$P(\nu)_{\pm}^{\text{min}} \leq P(\nu) \leq  P(\nu)_{\pm}^{\text{max}}$
if the mass hierarchy is known, where
the $+(-)$ sign refers to
$\Delta m^2_{31}>0$ ($\Delta m^2_{31}<0$)
and max (min) refers to the maximum (minimum) value of
the allowed region for
$P\equiv P(\nu_\mu\rightarrow\nu_e)$\@, respectively.
(See Fig.~\ref{envelope}.)
Without the knowledge of the mass hierarchy
the probability is within the region
$P(\nu)_{-}^{\text{min}} \leq P(\nu) \leq  P(\nu)_{+}^{\text{max}}$.
The similar inequalities are present also for
antineutrino appearance channel.
In Fig.~\ref{envelope} we present allowed regions
in the cases of $\Delta m^2_{31}>0$ and  $\Delta m^2_{31}<0$
on a plane spanned by
$P(\nu_{\mu} \rightarrow \nu_{e})$ and
$P(\bar{\nu}_{\mu} \rightarrow \bar{\nu}_{e})$ by taking two
best fit values $\sin^22\theta_{13} = 0.08,~0.04$
(labeled as a, b) as reactor affirmative cases.
They are inside the sensitivity region of the reactor experiment
discussed in section III\@.
We have used the one dimensional $\chi^2$
analysis (i.e., the only parameter is $\sin^22\theta_{13}$) to
obtain the allowed regions in Fig.~\ref{envelope}.
In doing this we have used the same systematic error of $0.8\,\%$ and
the statistical errors corresponding to \mbox{40\,ton-year} measurement
by the detector considered in section III\@.
For $\sin^22\theta_{13} \alt 0.02$,
the particular reactor experiment would fail
(the case of Reactor Negative) but the allowed region
can be obtained by the same procedure, and presented in
Fig.~\ref{envelope}, the region labeled as c.
We use the same LMA parameters as used earlier for Fig.~\ref{degen1}
and Fig.~\ref{degen2}.

We discuss four cases depending upon the two possibilities of
affirmative and negative evidences (denoted as Affirmative, and Negative)
in each disappearance and appearance search in reactor and
long-baseline accelerator experiments, respectively.
However, it is convenient to organize our discussion by classifying
them into two categories, (Reactor Affirmative), and (Reactor Negative).

\subsection{Reactor Affirmative}

We have two alternative cases, the LBL appearance
search Affirmative, or Negative.

\noindent
{\bf LBL Affirmative}:

Implications of affirmative evidence in the appearance search in LBL
experiments differ depending upon which region
the observed appearance probability $P(\nu)$ falls in:

\noindent
(1) $P_{-}^{\text{min}} \leq P(\nu) \leq P_{+}^{\text{min}}$, or
(2) $P_{-}^{\text{max}} \leq P(\nu) \leq P_{+}^{\text{max}}$:

These cases correspond to the two intervals which are given by
the projection on the $P$ axis of all the shadowed regions
(a or b)
minus the projection on the $P$ axis of the darker
shadowed region (a or b) in Fig.~\ref{envelope}.
It is remarkable that in these cases not only the sign of
$\Delta m^2_{31}$ is determined, but also the CP phase $\delta$
is known to be nonvanishing.
If $P(\nu)$ is in the former region then
$\Delta m^2_{31}$ is
negative and $\sin\delta$ is positive, whereas
if $P(\nu)$ is in the latter then
$\Delta m^2_{31}$ is
positive and $\sin\delta$ is
negative.

\noindent
(3) $P_{+}^{\text{min}} \leq P(\nu) \leq P_{-}^{\text{max}}$:

This case corresponds to the interval which is given by
the projection on the $P$ axis of the darker
shadowed region (a or b) in Fig.~\ref{envelope}.
In this case, neither
the sign of $\Delta m^2_{31}$ nor the sign of $\sin\delta$
can be determined.

It may be worth noting that if the reactor determination of
$\theta_{13}$ is accurate enough, it could be advantageous
for LBL appearance experiments to run only in the neutrino mode
(where the cross section is larger than
that for antineutrinos by a factor of 2-3)
to possibly determine the sign of $\Delta m^2_{31}$
depending upon which region $P(\nu)$ falls in.

\noindent
{\bf LBL Negative}:

In principle, it is possible to have no appearance event even
though the reactor sees evidence for disappearance.
This case corresponds to the left edge of the analogous
shadowed region in the case of $\sin^22\theta_{13}\simeq0.02$
in Fig.~\ref{envelope},
i.e., the allowed region with $\sin^22\theta_{13}\simeq0.02$
for which $P_{-}^{\text{min}}$
on the $P$ axis falls below $P=0.005$.
In order for this case to occur the sensitivity limits
$P(\nu)_{\text{limit}}$ of the LBL experiment must satisfy
$P_{-}^{\text{min}} <  P(\nu)_{\text{limit}}$
assuming our ignorance to the sign of $\Delta m^2_{31}$.
If it occurs that
$P_{-}^{\text{min}} < P(\nu)_{\text{limit}} < P_{+}^{\text{min}}$,
then the sign of $\Delta m^2_{31}$ is determined to be minus.

$P(\nu)_{\text{limit}}$ of the JHF experiment in its phase I
is estimated to be $3 \times 10^{-3}$
\cite{JHF}.\footnote{
%%%%%%%%%%%%%%% footnote %%%%%%%%%%%%%%%%%%
The sensitivity limit of $\sin^2{2 \theta_{13}}$ quoted in
\cite{JHF}, $\sin^2{2 \theta_{13}} \leq 6 \times 10^{-3}$,
obtained by using one-mass scale approximation ($\epsilon \ll 1$)
may be translated into this limit for $P(\nu)$.
}
Therefore, by using the mixing parameters typical to
the LMA solution, the case of LBL Negative cannot occur
unless the sensitivity of the reactor experiment becomes
$\sin^2{2 \theta_{13}} \alt 0.01$.
However, in the intermediate stage of the JHF experiment,
where $P(\nu)_{\text{limit}}$ is larger than $3 \times 10^{-3}$,
this situation may occur.

\subsection{Reactor Negative}

If the reactor experiment does not see disappearance of $\bar{\nu}_e$
one obtains the bound $\theta_{13} \leq \theta_{13}^{\text{RL}}$.
We have again two alternative cases, the LBL appearance
search Affirmative, or Negative.

\noindent
{\bf LBL Affirmative}:

If a LBL experiment measures the oscillation probability $P(\nu)$.
Then, for a given value of $P(\nu)$ the allowed region of
$\sin{2 \theta_{13}}$ is given by
$\sin{2 \theta_{\pm}^{\text{min}}} \leq
\sin{2 \theta_{13}} \leq
\sin{2 \theta_{\pm}^{\text{max}}}$
if the sign of $\Delta m^2_{31}$ is known, and by
$\sin{2 \theta_{+}^{\text{min}}} \leq
\sin{2 \theta_{13}} \leq
\sin{2 \theta_{-}^{\text{max}}}$
otherwise.
We denote below the maximum and the minimum values of $\theta_{13}$
collectively as $\theta_{\text{max}}$ and $\theta_{\text{min}}$, respectively.
In Fig.~\ref{degen2}, the region bounded by
$\sin{2 \theta_{+}^{\text{min}}}$ and $\sin{2 \theta_{+}^{\text{max}}}$
($\sin{2 \theta_{-}^{\text{min}}}$ and $\sin{2 \theta_{-}^{\text{max}}}$)
are indicated as a region bounded by the solid (dashed) black line
for a given value of $s^2_{23}$.

Then, there are two possibilities which we discuss one by one:

\noindent
(i) $\theta_{13}^{\text{RL}} \geq \theta_{\text{max}}$:
In this case no additional information is obtained by nonobservation
of disappearance of $\bar{\nu}_e$ in reactor experiment.

\noindent
(ii) $\theta_{\text{min}} \leq \theta_{13}^{\text{RL}} \leq \theta_{\text{max}}$:
In this case we have a nontrivial constraint
$\theta_{\text{min}} \leq \theta_{13} \leq \theta_{13}^{\text{RL}}$.

\noindent
{\bf LBL Negative}:

In this case, we obtain the upper bound on $\theta_{13}$, which
however depends on the assumed values of $\delta$ and the sign of
$\Delta m^2_{31}$. A $\delta$-independent bound can also be derived:
$\theta_{13} \leq$ min[$\theta^{\text{RL}}, \theta_{\text{max}}$].

\section{Discussion and Conclusion}

In this paper, we have explored in detail the possibility
of measuring $\sin^22\theta_{13}$ using reactor neutrinos.
We stressed that this measurement is free from
the problem of parameter degeneracies from which accelerator
appearance experiments suffer, and that the reactor measurement
is complementary to accelerator experiments.
We have shown that the sensitivity to
$\sin^22\theta_{13}\agt0.02$ (0.05)
is obtained with a $24.3\,{\text{GW}_{\text{th}}}$ reactor
with identical detectors at near and far distances and with
data size of \mbox{40 (10)\,ton-year}
assuming that the relative systematic error is
\mbox{0.8\,\%} (\mbox{2\,\%}) for the total number of events.
In particular,
if the relative systematic error is \mbox{0.8\,\%}, the
error in $\sin^22\theta_{13}$ is 0.018 which is
smaller than the uncertainty due to the combined
(intrinsic and hierarchical) parameter degeneracies
expected in accelerator experiments.
We also have shown that the reactor measurement can
resolve the degeneracy in $\theta_{23}\leftrightarrow\pi/2-\theta_{23}$
and determine whether $\theta_{23}$ is smaller or larger
than $\pi/4$ if $\sin^22\theta_{13}$ and $\cos^22\theta_{23}$
are relatively large.

We have taken \mbox{2\,\%} and \mbox{0.8\,\%} as the reference values for
the relative systematic error for the total number of events.
\mbox{2\,\%} is exactly the same figure as the Bugey experiment
while \mbox{0.8\,\%} is what we naively expect in the case we have
two identical detectors, near and far,
which are similar to that of the CHOOZ experiment.
It is also technically possible to dig a \mbox{200\,m} depth shaft hole with
diameter wide enough to place a CHOOZ-like detector in.
Therefore, the discussions in this paper are realistic.
We hope the present
paper stimulates interest of the community in reactor measurements
of $\theta_{13}$.

%%%%%%%%%%%%%%%% acknowledgments %%%%%%%%%%%%%%%%
\begin{acknowledgments}
We thank Yoshihisa Obayashi for correspondence and
Michael Shaevitz for useful comments.
HM thanks Andre de Gouvea for discussions, and
Theoretical Physics Department of Fermilab for hospitality.
HS thanks Professor Atsuto Suzuki and the members of
Research Center for Neutrino Science, Tohoku University,
for hospitality where core part of the sensitivity analysis
was carried out.
This work was supported by the Grant-in-Aid for Scientific Research
in Priority Areas No. 12047222 and No. 13640295, Japan Ministry
of Education, Culture, Sports, Science, and Technology.
\end{acknowledgments}

%%%%%%%%%%%%%%%%%%%%%%%%%% Bibliography %%%%%%%%%%%%%%%%%%%%%%%%%%%%%%

\newpage
%\vglue 4.5cm
\hglue -1.8cm
%\tightenlines
\begin{table}
\vglue 4.5cm
\hglue -1.8cm
\begin{tabular}{|l|c|c|c|}
\hline
Bugey & absolute normalization & relative normalization
& relative/absolute\\
\hline
flux& 2.8\% & 0.0\% & 0 \\
number of protons& 1.9\% & 0.6\% & 0.32 \\
solid angle& 0.5\% & 0.5\% & 1 \\
detection efficiency& 3.5\% & 1.7\% & 0.49 \\
\hline\hline
total& 4.9\% & 2.0\% & \\
\hline
\end{tabular}
%\vglue -2.63cm
\vglue 1.63cm
\hglue -0.2cm
%\tightenlines
\begin{tabular}{|l|c|c|c|}
\hline
CHOOZ--like & absolute normalization & relative normalization (expected)
& relative/absolute\\
\hline
flux& 2.1\% & 0.0\% & 0 \\
number of protons& 0.8\% & 0.3\% & 0.38 \\
detection efficiency& 1.5\% & 0.7\% & 0.47 \\
\hline\hline
total& 2.7\% & 0.8\% & \\
\hline
for bins& 8.1\% & 2.4\% & \\
\hline
\end{tabular}
\vglue 2.5cm
\label{tab:error}
\caption{Systematic errors in the Bugey and the CHOOZ-like experiments.
Relative errors in the CHOOZ-like experiment are expectation
with the same reduction rates of errors as those of Bugey.}
\label{table}
\end{table}
\newpage
%%%%%%%%%%%%%%%%%%%%%%%%%%%%%%%%%%%%%%%%%%%%%%%%%%%%%%%%%%%%%%%%%%%%%%
\vglue 4.cm
%\hglue 2.5cm

%%%%%%%%%%%%%%%%%%%%%%%%%%%%%%%%%%%%%%%%%%%%%%%%%%%%%%%%%%%%%%%%%%%%%%
\newpage

\newpage

%%%%%%%%%%%%%%%%%%%%%%%%%%%%%%%%%%%%%%%%%%%%%%%%%%%%%%%%%%%%%%%%%%%%%%

\newpage

%%%%%%%%%%%%%%%%%%%%%%%%%%%%%%%%%%%%%%%%%%%%%%%%%%%%%%%%%%%%%%%%%%%%%%

\newpage

%%%%%%%%%%%%%%%%%%%%%%%%%%%%%%%%%%%%%%%%%%%%%%%%%%%%%%%%%%%%%%%%%%%%%%

\newpage

%%%%%%%%%%%%%%%%%%%%%%%%%%%%%%%%%%%%%%%%%%%%%%%%%%%%%%%%%%%%%%%%%%%%%%
%%%%%%%%%%%%%%%%%%%%%%%%%%%%%%%%%%%%%%%%%%%%%%%%%%%%%%%%%%%%%%%%%%%%%%
\newpage

%%%%%%%%%%%%%%%%%%%%%%%%%%%%%%%%%%%%%%%%%%%%%%%%%%%%%%%%%%%%%%%%%%%%%%
\newpage

\end{document}